# Recent Progress and Future Prospects of 2D-based Photodetectors

*Nengjie Huo, Gerasimos Konstantatos\**

Dr. N. Huo, Prof. G. Konstantatos

ICFO – Institut de Ciencies Fotoniques, The Barcelona Institute of Science and Technology, Castelldefels, 08860, Barcelona, Spain
E-mail: Gerasimos.konstantatos@icfo.es

Prof. G. Konstantatos
ICREA – Institució Catalana de Recerca i Estudis Avançats, Lluis Companys 23, 08010 Barcelona, Spain

**Abstract:** Conventional semiconductors such as silicon and InGaAs based photodetectors have encountered a bottleneck in modern electronics and photonics in terms of spectral coverage, low resolution, non-transparency, non-flexibility and CMOS-incompatibility. New emerging 2D materials such as graphene, TMDs and their hybrid systems thereof, however, can circumvent all these issues benefitting from mechanically flexibility, extraordinary electronic and optical properties, as well as wafer-scale production and integration. Heterojunction-based photodiodes based on 2D materials offer ultrafast and broadband response from visible to far infrared range. Phototransistors based on 2D hybrid systems combined with other material platforms such as quantum dots, perovskites, organic materials, or plasmonic nanostructures yield ultrasensitive and broadband light detection capabilities. Notably the facile integration of 2D-photodetectors on silicon photonics or CMOS platforms paves the way towards high performance, low-cost, broadband sensing and imaging modalities.

## 1. Introduction

Photodetectors that convert light to electrical signals are one of key components in modern multifunctional technologies. So far, the developed photodetector technologies have covered the whole application spectrum that profoundly affects our daily lives: X-rays for biomedical imaging;[1] ultraviolet for lithography and living cell inspection;[2,3] visible light detection for digital camera and video imaging;[4] broad-range infrared detection for night vision, optical communications, atmospheric and quality inspection spectroscopy,[5-8] among others. Photodetectors are characterized with some key figures of merit, including responsivity, quantum efficiency, photo-gain, *etc*. In Box1, the relevant terminology, units and key figures of merits for photodetectors are summarized. To direct compare the performance among different devices, the detectivity $D^*$, in units of Jones (*i.e.* cm Hz$^{1/2}$ W$^{-1}$), is provided which takes into account variations in the detector geometry and noise measurement conditions. Today crystalline silicon photodetectors integrated with complementary metal-oxide-semiconductor (CMOS) technology have reached a high level of maturity and performance in digital cameras and optical sensing systems. Silicon has low dielectric constant which allows the low capacitance and fast operation (~ns), however the indirect bandgap (~1.1 eV) limits its high optical absorption and operation wavelength coverage within visible and near-wavelength infrared (NWIR) region.[9] Currently, infrared photodetectors based on epitaxial-grown InGaAs, InSb, HgCdTe and type-II superlattices have allowed sensing in the infrared spectrum beyond silicon´s reach.[9,10] These technologies have been well developed and commercialized with high sensitivity up to $10^{13}$ Jones and broad spectral coverage from NWIR to long-wavelength infrared (LWIR) region.[9] The single photon detectors such as single-photon avalanche photodiode, superconducting single-photon detectors, etc, have also been developed with high efficiency, low timing jitter and well photon-number resolution in NWIR region for specific applications such as quantum communication.[11] Despite the maturity of infrared photodetector technologies, there are still some major roadblocks for large scale deployment. For example, HgCdTe and other exotic semiconductors are usually fabricated with molecular beam epitaxy (MBE) or metal-organic chemical vapour deposition (MOCVD), techniques which sinificantly increase the manufacturing complexity and cost. These detectors also suffer from CMOS-incompatibility due to the large lattice mismatch and require cryogenic temperature of operation to obtain low noise and high sensitivity. Besides, the conventional bulky semiconductors are rigid, brittle and opaque, preventing their specific applications from bendable and flexible optoelectronics.

Presently, the diversity of photodetector applications are growing. Some factors in terms of CMOS-compatibility, low-cost manufacturing, transparency and flexibility, which are beyond the limits of conventional semiconductors, are needed for new emerging wearable and portable optoelectronics. Along this direction, two-dimensional (2D) materials, colloidal quantum dots (CQDs) and organic semiconductors have been demonstrated as appealing potential platforms. CQDs based photodetectors have exhibited high photoconductive gain in the range of $10^3$-$10^6$ and detectivity up to $10^{13}$ Jones from visible to the shortwave-IR range by tuning the size of CQDs.[12,13] Organic photodiodes were recently reported with excellent *LDR* of 110 dB, -3dB bandwidth of 11.4 MHz and detectivity up to $10^{14}$ Jones,[14] showing their potential in high performance photodetectors. In this review, we focus on the development of 2D materials based photodetectors, functionalization with CQDs, organic materials, etc, to improve the performance, as well as the potential challenges to further develop and commercialize these technologies.

Since the discovery of graphene in 2004, 2D atomic sheets including graphene, transition metal dichalcogenides (TMDs) and black phosphorus (BP), with intralayer covalent bonding and interlayer van der Waals (vdW) interaction have emerged as a unique and promising material family for photonics and optoelectronics in view of their appealing characteristics.[15-17] Their atomically thin feature enables high transparency and mechanical flexibility, offering the opportunity for bendable, flexible and conformal detectors.[18, 19-22]. Due to their flexible nature, strain engineering can be exploited to modulate the electronic and optical properties. Desai *et al*. observed an indirect to direct bandgap transition in bilayer WSe$_2$ through applied strain, which led to the improvment in photoluminescence by 35 times.[23] Recently, Ahn *et al*. also demonstrated a dramatic modulation of the band structure through strain engineering introduced directly during the growth process.[24] Strain engineering can also be exploited in bulk semiconductors to tune their optical and electrical properties. But, the maximum achievable tuning is eventually determined by the elastic limit of the material (~1% in most bulky semiconductors). The atomically thin thickness of 2D materials leads to a high 'stretchability' up to ~20% strain magnitudes, providing a larger room for poreprty tuning via strain.[25] When 2D materials, particular TMDs, are thinned to monolayer limit, quantum confinement effects lead to the indirect-to-direct bandgap transitions, which can be used to tune the band structure and absorption wavelength by varying the number of layers.[26] Due to the direct bandgap, the monolayer TMDs have strong light-matter interactions from the strongly bounded excitons.[27,28] High in-plane mobility, of 10,000 cm$^2$ V$^{-1}$ s$^{-1}$ reported in graphene and 100-500 cm$^2$ V$^{-1}$ s$^{-1}$ in TMDs, at room temperature, facilitates efficient photo-carrier extraction and leads to fast and sensitive detectors.[29-31] Another important advantage of 2D materials is the absence of surface dangling bonds due to the vdW interlayer interactions, enabling seamless integration on any substrate crystalline or amorphous, rigid or flexible.[32-34] Large area growth and easy-processing of 2D materials warrant low-cost and large-scale manufacturability.[35] Finally, the versatility of 2D family materials offer different applications ranging from semi-metal graphene, semiconducting TMDs to insulating h-BN and new monoatomic buckled crystals like black phosphorous (BP)[36] and silicene,[37], which can be synergistically combined to build hybrid platforms for multifunctional optoelectronics.[32,38]

Benifiting from these extraordinary properties, 2D material-based photodetectors hold promise to overcome the roadblocks of current conventional photodetector technologies operated at room temperature. On the other hand, 2D materials and their photodetectors also face some challenges. The ultrahigh mobility of graphene makes it suitable for high-speed photodetectors (up to 40 GHz bandwidth),[29] but its single atomic layer and zero bandgap limit its photo-absorption, external quantum efficiency and thus photoresponsivity. Graphene photodetectors are characterized by large dark current that is typically associated to high noise and power consumption. Various approaches have been proposed to introduce a bandgap in graphene, such as building graphene QD-like structures[39] or cutting graphene into nanoribbons.[40] However, these bandgap-opening methods degraded the electronic performance and the speed of graphene photodetectors compared to their pristine counterparts. 2D TMDs with sizable bandgap can overcome the disadvantages of graphene, however, the photoresponse wavelength range is limited from ultraviolet to near-IR, determined by their optical bandgap. Also the response speed of TMDs photodetectors is relatively slow due to significant trapping effects of photocarriers.[31] BP has an intermediate bandgap between graphene and TMDCs, thus covers a broad photoresponse spectrum from near-IR to mid-IR. BP photodetectors were also reported with high responsivity of 657 mA/W and fast speed up to 3 GHz,[41] yet its stability remains a main issue for this class of materials.

In the first section of this review, we briefly discuss the classification of photodetectors as photoconductors, photodiodes and phototransistors. Then, we review the state-of-the-art 2D crystal based photodetectors in terms of the underlying physical mechanisms with and without gain. We focus on the strategies developed recently to address the above-mentioned drawbacks and further improve the performance of 2D photodetectors via doping or sensitization with colloidal quantum dots (CQDs), perovskites, metal nanostructures and organic molecules or polymers. In the third section, we demonstrate current applications of 2D photodetectors on flexible substrates or integrated with silicon and CMOS technologies. In the last section, we compare the performance between commercial photodetectors and developed 2D photodetectors, and outline the new challenges needed to be addressed to further boost performance, functionality and commercialization of this new optoelectronic platform. These challenges include improvement in linear dynamic range, contact and mobility engineering, noise suppression issues and advances in wafer-scale manufacturing, among others.

**Box**

**Figures of merit:**

**Responsivity $R$**, Responsivity is defined as $R = \frac{I_{\text{ph}}}{P_{\text{in}}}$, representing the ratio of photocurrent generated in the detector over the incident optical power in units of A W$^{-1}$, where $I_{\text{ph}}$ is photocurrent, $P_{\text{in}}$ incident light power.

**External quantum efficiency**, *EQE* is the ratio of the number of photogenerated electrons in a detector over the number of incident photons and is given as $EQE = \frac{I_{\text{ph}}}{P_{\text{in}}}\frac{h\nu}{e} = R\frac{h\nu}{e}$

**Internal quantum efficiency**, *IQE* is the ratio of the number of photogenerated electrons in a detector over the number of absorbed photons and is given as $IQE = R\frac{h\nu}{A_a e}$, where $A_\alpha$ is the absorbed fraction of light.

**Linear dynamic range**, *LDR* is used to characterize the light intensity range in which the photodetectors have a constant responsivity in unit of dB and can be expressed as $LDR = 10 \times \log_{10}(\frac{P_{\text{sat}}}{NEP})$ [dB], where *NEP* is noise equivalent power (defined below), $P_{\text{sat}}$ the saturated light intensity at which the photocurrent begins to deviate from linearity.

**Signal to Noise Ratio**, *SNR* is the ratio of signal power to noise power, which is must be larger than unity so that the signal power can be distinguished from the noise.

**Photo-Gain, $G$** describes the number of recirculated carriers in the circuit per single incident photon. In a photodiode, the photo-gain is equal to unity, unless carrier multiplication effects are present. In a photoconductor, one type of carrier (e.g. hole) is usually captured in trap states or sensitizing centres with a lifetime of $\tau_{life}$, while the other type of carrier (e.g. electron) is free to traverse the channel with a transit time of $\tau_{transit}$. If carrier lifetime is longer than transit time, the free electrons recirculate many times before recombination with captured holes, leading to a generation of photo-gain, which is defined by $G = \frac{\tau_{life}}{\tau_{transit}} = \frac{\tau_{life}}{L^2}\mu V_{DS}$, where $L$ is length of channel, $\mu$ is carrier mobility, $V_{DS}$ is applied bias across channel.

**Noise equivalent power**, *NEP* is the minimum detectable optical power at which the electrical signal-to-noise ratio (*SNR*) in the detector is equal to unity, when bandwidth is limited to 1 Hz. The unit is W Hz$^{-1/2}$. *NEP* describes the sensitivity of a detector and is determined by both noise current and responsivity, $NEP = \frac{I_{noise}}{R}$.

**Detectivity *D*\***, is another common figure of merit to characterize the sensitivity of a detector, which can enable the comparison of detectors with different geometries. The unit is cm Hz$^{1/2}$ W$^{-1}$ or Jones. It can be defined by $D^* = \frac{\sqrt{AB}}{NEP} = \frac{R\sqrt{AB}}{I_{nosie}} = \frac{R\sqrt{A}}{S_n}$, where *A* is active area, *B* is electrical bandwidth, $S_n$ is noise spectral density.

## 2. Classification of photodetectors

### 2.1. Photodiodes

Typically, the two most common photodetector types are photodiodes and photoconductors. 2D crystals have been widely explored in both classes of detectors. In a photodiode, a built-in field is formed via a p-n junction or a Schottky junction between metal and semiconductors. The generated electrons and holes by incoming photons move to opposite contact electrodes driven by the built-in potential. The response speed is determined by the transit time of excess charge carriers defined by $t_{transit} = L^2/\mu V_{bi}$, where L is channel length, $\mu$ is carrier mobility and $V_{bi}$ is the built-in potential across junction.[42] For graphene based photodiodes, the mobility is over 10,000 cm$^2$ V$^{-1}$ s$^{-1}$ leading to ultrafast speed of picoseconds to nanoseconds and large bandwidth on the order of Gigahertz.[29,43-45] On the other hand, the quantum efficiency of photodiodes is limited to unity, unless avalanche or carrier multiplication effects occur when operated near the breakdown regime, offering the possibility of multiple carriers generated per single photon; this comes at the cost of very high bias of 50-100 V that is needed to reach this regime. 2D materials such as $MoS_2$, InSe and BP based avalanche photodiodes (APDs) have been demonstrated with carrier multiplication of 100-1000.[46-48]

## 2.2. Photoconductors

In a photoconductor, the photo-excited electrons and holes are separated by the applied external bias and drift to opposite Ohmic contact electrodes, leading to photocurrent or photo-voltage generation. A gain mechanism, called photoconductive gain, can be present in this case when one type of charge carriers is able to circulate through an external circuit many times before it recombines with its opposite carrier. The gain is then defined as the ratio of lifetime and transit time (see Box 1). Usually, the trap states or sensitizing centres within the band gap of the semiconductors capture and localize one charge carrier type and effectively prolong its carrier lifetime, leading to multiple carriers per single photon and extremely high responsivity compared to a photodiode.[42,49] Due to the fact that both gain and temporal response are determined by the lifetime of trapped carriers, the bandwidth in this class of detectors is usually lower than that of the photodiodes, thus a trade-off between gain and response should be considered when designing detectors for specific applications.

## 2.3. Phototransistors

The strategies to improve the performance of photodetectors are not only maximizing its electrical response to light in terms of gain, but also minimizing the noise in its electrical output. Low noise can be achieved in high quality photodiodes with suppressed generation-recombination noise and negligible shot noise contribution when operated in a photovoltaic mode; the absence of gain though limits the electrical response, giving rise to noise contributions from the read-out electronics. On the other hand, the Ohmic contacts in photoconductors leads to large dark current and thereby large shot and 1/f noise components, that however may compete with the presence of gain to reach high *SNR* ratio under specific conditions. To address the high noise issue in photoconductors as mentioned above, the concept of phototransistors, a special case of photoconductors with an additional gate terminal which is electrically isolated from the semiconductor channel by a thin dielectric, is demonstrated in 2D materials based photodetectors.[31,49] The applied gate $V_G$ can electronically modulate the carrier density by field-effect and switch off the dark current by operating the device in the depletion regime. Thus both photoconductive gain and low noise can be achieved leading to high sensitivity and gain-bandwidth products. As a typical example, monolayer $MoS_2$ with a direct band gap of 1.8 eV possesses large gate-tunable conductivity with on/off ratio exceeding $10^8$ and extremely low off current densities of 25 fA $\mu m^{-1}$,[30,50] resulting to reported responsivities of ~1000 A $W^{-1}$ and experimentally measured detectivity of $10^{11}$-$10^{12}$ Jones.[31,50-52]

**3. 2D-based photodetectors**

In this section, we briefly discuss the fundamental properties of 2D crystals and the progress of 2D based photodetectors.

**3.1. Crystal structure and basic properties**

2D crystal systems encompass a large number of family members with unique and varying physical properties, including metals, semimetals, semiconductors, insulators. This enables the development of an entirely independent new generation of optoelectronics based exclusively on 2D materials. Depending on the specific application in terms of sensitivity, speed or wavelength spectrum coverage, different 2D materials are selected and designed. The chemical structure of various monolayer 2D materials including semi-metal graphene, semiconducting TMDs, insulating h-BN and monoatomic buckled crystals are shown in **Figure 1**. Graphene, monolayer carbon atoms bonded together in a hexagonal honeycomb lattice, is a gapless semi-metal material with ultrahigh mobility,[53-55] and as such an appealing material for broadband photodetection from terahertz (THz) to ultraviolet (UV)[56-58] and ultrafast technologies.[43-45] Graphene can also serve as work function tunable electrode due to linear dispersion of the Dirac electrons near the K point.[59-61] 2D TMDs with a chemical formula $MX_2$ (M: Mo, W, *etc*. and X: S, Se or Te), where each layer is composed of one layer of hexagonally packed metal atoms sandwiched with two layers of chalcogen atoms, have a variable bandgap of 1-2 eV and offer strong light-mater interactions,[62,63] which can be utilized as field effect channel semiconductors and photoactive layers for strong light absorption. 2D h-BN with an ultrawide bandgap of ~6 eV, where boron and nitrogen atoms are covalently bonded in each layer, is an excellent insulator as high-ƙ dielectric to enhance device performance such as mobility and stability.[64,65] Recently, new monoatomic buckled crystals (termed Xenes, Figure 1) like black phosphorene,[36,66,67] silicene,[37,68,69] germanene,[70,71] and bismuthene,[72,73] *etc*. have been theoretically predicted and experimentally developed, which can offer smaller bandgap of 0.2-2 eV, high mobility of 100-1000 $cm^2$ $V^{-1}$ $s^{-1}$ and the possibility to serve as high performance short- and mid-infrared photodetectors.[74-76] All these 2D crystals share the same advantages of transparency, flexibility, vdW interlayer interactions and CMOS compatibility. The well-developed manufacturing methods such as mechanical/liquid phase exfoliation,[77,78] CVD growth[35,79-81] and inkjet-printing[82,83] enable large-scale fabrication and easy-processing for both fundamental research and practical applications in optoelectronics.

## 3.2. 2D-based photodetectors without gain

In this section, we discuss the 2D materials based photodetectors without gain and the corresponding photo-detection mechanisms. Typically, this class of photodetectors include 2D-photodetectors with photo-thermoelectric and bolometric effects and 2D heterojunction-based photodiodes based on the photovoltaic effect.

### 3.2.1. Photo-thermoelectric and bolometric detectors

The photo-thermoelectric effect (PTE) refers to the carriers generated by the photo-induced temperature gradient $\triangle T$ between different substances upon localized illumination. Thus a photovoltage $V_{PTE}$ can be produced through the Seebeck effect (or thermoelectric effect): $V_{PTE} = (S_1-S_2) \triangle T$, where $S_1$ and $S_2$ (in V K$^{-1}$) are the Seebeck coefficients of the two materials (**Figure 2**a). The photovoltage can drive the hot carriers to form a net current flowing without external bias. Due to the dominant hot carrier transport and strong e-e interactions in graphene, the PTE effect plays a key role in the photo-response of graphene p-n junctions.[84-89] In some reported metal-graphene-metal photodetectors, the photocurrent is dominated by the photovoltaic effect.[90-92] Graphene photo-thermoelectric detectors exhibit high bandwidth up to 40 GHz, the responsivity however, is limited to 6.1 mA W$^{-1}$ due to the low optical absorption and the absence of gain.[29,44,45,86-89] Beside graphene, significant PTE effect has also been reported in MoS$_2$ detectors arising from the large mismatch of Seebeck coefficients between MoS$_2$ and metal electrodes.[93-95] The Seebeck coefficient of MoS$_2$ (30-100 mV K$^{-1}$) is several orders of magnitude larger than that of graphene (4-100 μV K$^{-1}$).[93-95] Black phosphorene[96,97] and 2D SnS$_2$[98] have also been reported with large Seebeck coefficients of 60-335 μV K$^{-1}$ and 34.7 mV K$^{-1}$, respectively, rendering them promising candidates in thermoelectric detectors.

The bolometers, on the other hand, are used to detect the changes in incident photon radiation (d$P$) by measuring the changes in temperature (d$T$) of the absorbing element. The bolometric effect is associated with the conductance change (d$G$) of a semiconductor channel induced by photo-induced heating of an absorber and thus an external bias is needed to monitor the d$G$ which is followed by the d$T$-induced change of the mobility (Figure 2b).[99,100] The sensitivity of bolometers is ultimately determined by the thermal conductance: $G_h$ = d$P$/d$T$. High-sensitivity bolometers, which are made of doped or ion-implanted semiconductors,[101] show a strong temperature-dependence of its conductance at operating temperature upon illumination in the far-infrared and submillimetre (THz) wavelength range. Upon light irradiation, the temperature changes as a function of time with thermal time constant $\tau = C_h/G_h$, where $C_h$ is the heat capacity which determines the response time.[99] The bilayer graphene-based bolometers have exhibited very fast intrinsic speeds (>1 GHz at 10 K) and high sensitivity (33 fW Hz$^{-1/2}$ at 5 K) due to the small heat capacity and weak electron-phonon coupling.[102] Very recently, Wu *et al.* demonstrated a negative infared photoresponse (up to 1550 nm) in multilayer MoS$_2$ for the first time with incident photon energy lower than bangap of MoS$_2$. The responsivity at 980 nm can reach 2.3 A W$^{-1}$ with response time of 50 ms. The negative photoresponse was attributed to the bolometric effect with a bolometric coefficient of -33 nA K$^{-1}$. The increased temperature from light irradiation could lead to the decreased conductivity of MoS$_2$ due to the increased electron-phonon interaction.[103] In bolometers, the temperature coefficient of resistance (TCR in units of % K$^{-1}$) is another key performance indicator and defined as: $\text{TCR}(R_0) = \frac{1}{R_0}\frac{dR}{dT}$, representing the percentage change in resistance per Kelvin at the operating resistance $R_0$. Recently, graphene-based pyroelectric bolometers have reached a record TCR up to 900% K$^{-1}$ at room temperature and temperature resolution down to 15 μK by incorporating a pyroelectric substrate (LiNbO$_3$) and a floating metallic structure (Figure 2c).[104] Goswami *et al.* demonstrated a TCR up to -2.9% K$^{-1}$ at room temperature in pulsed laser deposition (PLD) grown MoS$_2$ film on silicon substrate. Their reported MoS$_2$ bolometers were sensitive to mid-IR irradiation (7-8.2 μm) with responsivity of 8.7 V W$^{-1}$ and response time on the order of 10 s.[105] Superconductors are also known to show a very strong resistance dependence on temperature near the critical temperature $T_c$, and have been proposed as sensitive superconducting bolometers.[99] Recently, many 2D materials such as MoS$_2$,[106] Mo$_2$C[107] and FeSe[108] are reported with remarkable superconductive properties possessing high critical temperature, thus they may serve as potential new material platforms for 2D superconducting bolometers due to their extremely sensitive resistance change upon temperature.

### 3.2.2. 2D heterojunction based photodiodes

Van der Waals heterojunctions designed by assembling isolated 2D crystals have emerged as a new class of artificial materials with extraordinary optoelectronic functionalities in various applications such as solar cells, photodetectors and light emitting diodes. The 2D heterojunctions are fabricated by either local chemically or electrostatically induced doping or through transfer technologies in both in-plane and out-of-plane directions. These heterojunctions have been typically explored as photodiodes, where ultrafast response is possible followed by the absence of gain. The underlying physical mechanism of photo-detection is the photovoltaic effect based on the separation of photo-generated electron–hole (e–h) pairs by the built-in electric field at the junction. Next, we discuss the progress in heterojunction-based photodiodes with both in-plane and out-of-plane configurations.

*In-plane heterojunctions:* The atomically thin profile of 2D crystals enables the efficient modulation of their carrier density, work function and polarity through electrostatic doping[36,109] or chemical doping.[110,111] As a result, a large number of in-plane heterojunction based photodiodes have been fabricated. One class of in-plane photodiodes is realized by the localized electrostatic doping on 2D semiconducting channels including graphene,[53] $WSe_2$[109] and BP[36] with natural ambipolar transport behaviour. **Figure 3**a shows a typical device structure of $WSe_2$ based lateral diode, where the split gate electrodes are located under a high-ƙ dielectric such as $HfO_2$, h-BN or SiN.[112-115] The local gate can modulate the carrier type and concentration in the $WSe_2$ channel atop. Electrons or holes can be induced respectively by applying opposite polarity bias on the split gate electrodes, thus forming a lateral P-N junction in a single $WSe_2$ channel. In this case, the device can act as photodiode with evident rectifying behaviour under dark and photovoltaic effect under illumination, leading to the generation of short circuit current $I_{SC}$ and open circuit voltage $V_{OC}$ (Figure 3b). The responsivity of the in-plane $WSe_2$ photodiode has been reported from 0.7-210 mA $W^{-1}$ with tens of milliseconds response time. The EQE is also limited to 0.1%-0.2% due to the weak photon absorption and finite depletion width.[113,116] In the same device configuration and operation principle, BP[117] and MoSe$_2$[118] have also been demonstrated as in-plane photodiodes with split gate well-controlled photocurrent generation and transport. Due to the narrow bandgap of BP, the BP photodiodes have an extended spectrum to NIR region (up to 1500 nm) with responsivity of 28 mA W$^{-1}$, response time constant of 2 ms and EQE of 0.1%.[117]

Another class of in-plane photodiodes has been developed via doping or direct CVD-growth. P-type doping techniques for MoS$_2$ have been reported recently including plasma treatment,[119] niobium (Nb) physical doping[120] and gold chloride (AuCl$_3$) chemical doping.[121] In a MoS$_2$ phototransistor, one can use an insulator such as h-NB or HfO$_2$ to cover partially the MoS$_2$ channel protecting it from doping the underneath layer (Figure 3c), thus a lateral PN junction is formed at the interface to perform rectifying and photovoltaic properties with responsivity of 300 mA W$^{-1}$.[122] The same device concept was also explored in BP[123] where benzyl viologen (BV) was used as electron dopant; the obtained responsivity was 180 mA W$^{-1}$ at a wavelength of 1.5 µm. Recently, many groups[124-126] have developed the one- or two-step epitaxial growth of lateral TMD-TMD heterostructures, implemented in WSe$_2$-MoS$_2$ and WS$_2$-WSe$_2$ heterojunctions that exhibited atomically sharp interfaces (Figure 3d). Sahoo *et al*. developed a one-pot synthetic approach, using a single heterogeneous solid source, for the continuous fabrication of lateral multi-junction heterostructures consisting of monolayers of varing TMDs (Figure 3e). They also fabricated the heterojunction based devices (Figure 3f) and observed significant photocurrent across the junctions (Figure 3g).[127] These results enabled continuous growth of multi-junction lateral heterostructures and realization of lateral p-n diodes and photodiodes as well as complex in-plane superlattices. Thanks to their higher quality interface over previous architectures this class of photodide have reached EQE of 9.9% (corresponding to an IQE of 43%) and fast response speed of 100 µs.[126]

*Out-of-plane heterojunctions:* Heterojunctions based on 2D materials can have another form factor in that they may form instead across the van der Waals interfaces or different 2D crystals, also known as out-of-plane heterojunctions. We summarize them herein as TMD-TMD vertical heterojunctions and Gr-TMD-Gr sandwich structures. For the former case, benefiting from the well-developed transfer technologies, the atomically sharp surfaces and the absence of lattice-match constraints, a variety of 2D TMDs are stacked on each other to form high quality vertical heterojunctions.[128-132] As an example, $MoS_2$ and $WSe_2$ possess opposite intrinsic N-type and P-type behaviours, respectively, thus the stacking of those can form vertical PN junctions which have been extensively demonstrated with very promising photovoltaic and electroluminescent properties paving the way towards all-2D photovoltaics, photodetectors and light emitters.[129,133] Other TMDs have also been combined to exhibit diode-like behaviour and efficient photocurrent generation due to the type-II band alignment that can facilitate efficient electron–hole separation for light detection and harvesting.[134] For example, Hong, *et al*. reported ultrafast (~50 fs) charge transfer at the $MoS_2/WS_2$ interface because of the type-II band alignment, showcasing the large potential of vertical 2D heterojunctions in optoelectronic applications that require very high speed including ultra-fast spectroscopy and optical, on-chip communications.[135] Among the widely reported vertical heterojunction-based photodiodes, the responsivity is on the order of 10 mA/W, with reported EQE up to 80% and response time in the millisecond range.[128-132] The higher quantum efficiency compared to that of in-plane photodiodes is attributed to the larger photoactive area (stacking region), strong interlayer coupling and efficient interlayer charge transfer. In such heterojunstions, some new physical mechanisms emerge resulting in drastic improvments in device performance. Barati *et al.* demonstrated a monolayer $MoSe_2$ and bilayer $WSe_2$ heterojunction based photocells (**Figure 4a**) and observed highly efficient multiplication of interlayer e-h pairs which are generated by hot-electron impact excitation at temperature near 300 K. As shown in Figure 4b, the reverse bias photocurrent increases rapidly with increasing temperature indicating that an optically excited electron in the conduction band of $WSe_2$ undergoes efficient multiplication of electrons in $MoSe_2$. Such interlayer impact excitation results in responsivity enhancement exceeding 350% with a multiplication factor over 3.5 at low irradiance.[136] In addition to TMD-TMD junctions, other 2D materials were also combined with TMDs to form photodiodes. Such architectures have been based on the combination of TMDs with graphene (or BP), where broadband detection up to 2.4 µm (or 1.5 µm) range with microsecond response speed and high $D^*$ of $10^{11}$ Jones were achieved.[137-139] Recently, Long *et al.* reported a black arsenic phosphorus (b-AsP)-based long-wavelength IR photodetectors with room temperature operation up to 8.2 µm. By combing with multilayer $MoS_2$ (Insert of Figure 4c), the b-AsP-$MoS_2$ heterojunction based devices showed a diode behaviour (Figure 4c), and the photodiode can exhibit a specific detectivity higher than $4.9 \times 10^9$ Jones in 3-5 µm range, high responsivity of 115.4 mA $W^{-1}$ at 4.29 µm and fast speed of < 0.5 ms as well as suppressed flicker noise; these values are well beyond all room temperature Mid-IR photodetectors to date (Figure 4d).[140]

Now we turn to the Gr-TMD-Gr sandwich structures for photo-detection applications. This device concept (Figure 4e), where graphene serves as top and bottom transparent electrodes for charge extraction and the TMD acts as the photoactive layer, responsible for photon-absorption and carrier transport, is proposed based on the efficient gate-tunable Fermi level of graphene[141] and strong light-matter interactions of TMDs.[62,142] The first report of this class of detectors was a Gr-$WS_2$-Gr structure which acted as tunnelling transistors under dark.[142] A built-in field across $WS_2$ was created due to the different Fermi level position in graphene induced by electrostatic gating (Figure 4f). Under light illumination, the photo-excited electrons and holes are subsequently separated and drifted towards opposite graphene electrodes. The built-in field and thus photocurrent can be efficiently modulated by varying the gate voltage. The devices reached remarkable performance with responsivity of ~0.1 A $W^{-1}$ and EQE above 30%, demonstrated also as efficient flexible photovoltaic devices.[142] Then Yu *et al*. utilized multilayer $MoS_2$, instead of $WS_2$, to demonstrate gate-tunable photocurrent with EQE as high as 55% (IQE of 85%), fast response of 50 µs and responsivity of ~0.2 A $W^{-1}$ as shown in Figure 4g and 4h.[143] Similarly, M. Massicotte, *et al*. fabricated Gr-$WSe_2$-Gr heterostructures and reached unprecedented photo-response time as short as 5.5 ps combined with high EQE of 7.3% (IQE of 85%), that can be tuned by applying a bias and by varying the $WSe_2$ thickness, entering the regime of ultrafast photodetectors.[144] All these results indicate that the Gr-TMD-Gr sandwich structures in out-of-plane configuration allow highly efficient photo-detection capabilities owing to the strong light absorption in TMDs, atomically thin carrier transient paths for efficient charge separation and collection as well as practically sizeable junction areas for efficient photon harvesting. In this configuration, large room for improvement remains in particular towards broader spectral coverage employing 2D materials with smaller bandgap such as BP[36] and $Bi_2Se_3$.[145]

**3.3. 2D-hybrid photodetectors with Gain**

Although high quantum efficiency and ultrafast response were achieved in 2D heterojunction based photodiodes with both in-plane and out-of-plane configurations, the responsivity of those has been limited to 1 A $W^{-1}$ determined by the ceiling of unity quantum efficiency that photodiodes offer and the absence of gain. In view of this a new class of photodetectors have been proposed and developed based on 2D materials that offer the possibility of gain and the achievement of very high responsivity. These detectors rely on the photoconductive effect, that yields photoconductive gain due to the difference between the transit time of majority-type of carriers (e.g. electrons) and the lifetime of the minority carriers (e.g. holes), in which electrons recirculate multiple times before recombination with holes. The ratio of the carrier lifetime over the transit time determines the value of gain that can be reached. To demonstrate and even enhance the presence of photo-gain effects in 2D-based photodetectors, several approaches have been reported based on surface doping,[146] sensitizations with QDs,[147] perovskite[148,149] or metal nanostructures.[150] The common aim of those efforts has been to introduce or enhance the photogating effect. In the photogating effect, the minority carrier lifetime is prolonged trough localized trapping in trap states or sensitizing centers, that can be formed either on the 2D semiconductors themselves or via using sensitizer layers based on other material platforms.

The presence of such high gain comes at the expense of lower electrical bandwidth typically reported on the order of 1 kHz or lower. This feature determines then the employment of these detectors in applications that require high sensitivity but not very fast operation such as video imaging, sensing and steady-state spectroscopy applications. In this session, we discuss the different strategies developed so far to explore high photoconductive gain utilizing the photogating effect in 2D-hybrid photodetectors.

**3.3.1. Photogating effect in graphene sensitized detectors**

Graphene has demonstrated its potential as photodetecting material in broadband and ultrafast technologies, however the monolayer thickness of graphene absorbs only 2.3 % of incident light.[57,58] The dominant photo-thermoelectric or bolometric effects in photo-response and fast photo-carrier recombination rates of a few picoseconds have limited the responsivity and sensitivity in graphene photodetectors due to the absence of gain mechanism and the limited absorption. To enhance the photon absorption, several approaches have been demonstrated such as the integration of optical microcavities,[151,152] optical waveguides[45] and the field enhancement by plasmons,[153] which have led to improved absorption of more than 60% and responsivity in the 20-130 mA W$^{-1}$ range. Interested readers can see recent review focussing on this approach.[100] Here we discuss the graphene hybrid phototransistors sensitized with other 2D materials, colloidal quantum dots, perovskites and organic materials, *etc*, which can further improve the performance through the photogating effect.

*Graphene-CQDs or perovskite sensitized detectors:* The first report of a highly sensitive graphene-based photodetector was based on sensitization with colloidal quantum dots (CQDs) of PbS.[147] CQDs have many unique properties that make them an ideal sensitizing pair for such as strong light absorption, broad absorption range from ultraviolet to SWIR and size-tunable bandgap through quantum confinement effect.[12] Their low-temperature and facile solution processing make them compatible with various substrates and large-scale, low-cost manufacturing processes. The concept of hybrid graphene-CQD phototransistors is shown in **Figure 5**a. The detector consists of monolayer or bilayer graphene covered with a thin film of colloidal quantum dots. Upon light illumination, the photo-excited holes can transfer to graphene and drift by means of a voltage bias to the drain, while electrons remain trapped in the quantum-dot layer. These trapped carriers lead to a photogating effect, where the presence of these charges changes the conductance of graphene, and shifts the Dirac point to higher back-gate voltage (Figure 5b). Thanks to the high mobility of graphene and strong light absorption of CQDs, the hybrid phototransistors exhibited high EQE of 25%, high gain up to $10^8$ corresponding to responsivity of $10^7$ A W$^{-1}$ in the shortwave-IR region up to 1.6 µm determined by the size of the CQDs.[147] The high quantum efficiency, fast video-imaging speed (<10ms) and high detectivity of $10^{13}$ Jones, taken together with its CMOS compatibility rendered this hybrid systems a promising platform for visible and SWIR photodetection applications. Nikitskiy *et al.* then further integrated an electrically active colloidal quantum dot photodiode atop a graphene phototransistor (Figure 5c). By applying a bias $V_{TD}$ across the CQDs film, the additional electrical field perpendicular to the graphene-CQDs interface increases the depletion width and charge collection efficiency. As a result, the EQE was improved in excess of 70% (Figure 5d) with a linear dynamic range of 110 dB and 3 dB bandwidth of 1.5 kHz.[154]

Alternative sensitizers have then been employed in such architectures. As an example, topological insulator $Bi_2Te_3$ nanoplates were decorated on graphene and the responsivity of 35 A W$^{-1}$ (gain of ~83) with response spectrum from visible to near-infrared (980 nm) and telecommunication band (1550 nm) has been achieved in view of the small bandgap of $Bi_2Te_3$.[155] Large bandgap quantum dots like CdS and ZnO were also integrated with graphene to form hybrid phototransistors, showcasing promising applications in UV detectors and image sensors with high sensitivity and gain of $10^7$-$10^9$.[156,157] In a recent review,[158] various transition metal oxide (TMO) nanoparticles were also used in conjunction with graphene for different applications. Lee *et al*. reported graphene-perovskite ($CH_3NH_3PbI_3$) hybrids phototransistors, which exhibited a responsivity of 180 A W$^{-1}$ in the UV-Visible range, limited by the absorption range of the perovskite layer.[159] Using synergistically plasmonic effects of metallic nanostructures, the performance in this hybrid system was two-fold improved.[160] Recently, graphene-MAPbI$_3$ perovskite hybrid detectors have shown further progress with responsivity of $10^7$ A W$^{-1}$ and detectivity up to ~$10^{15}$ Jones, thanks to the efficient light harvesting from perovskite sensitizers, its very long carrier diffusion length and the efficient charge separation at the interface.[161]

***Graphene-2D TMDs sensitized detectors:*** The strong optical absorption ($1 \times 10^7$ m$^{-1}$) and a visible-NIR range bandgap of 2D TMDs, make them natural partners of graphene for optically active heterostructures.[142] Roy *et al.* fabricated the graphene-on-MoS$_2$ binary heterostructures and showed remarkable dual optoelectronic functionality, including highly sensitive photodetection and gate-tunable persistent photoconductivity.[162] The schematic diagram of the device is shown in Figure 5e, where the multilayer MoS$_2$ acting as the photoactive layer lies underneath the graphene and source-drain electrodes located atop render graphene as the carrier transport layer. Under light illumination, the photo-generated electrons in MoS$_2$ transferred into graphene under negative gate bias (Figure 5f) changing the conductance of graphene, while holes were trapped by localized states in the MoS$_2$ acting as a local gate and giving rise to the photogating effect. The ultrahigh mobility of graphene ($10^4$ cm$^2$ V$^{-1}$ s$^{-1}$) allowed the fast transit time of electrons, whereas holes residing longer time in MoS$_2$ result in a very large photoconductive gain of ~$4 \times 10^{10}$ electrons per single photon and reported responsivity of $5 \times 10^8$ A W$^{-1}$ in the visible at room temperature. At the expense of this gain, the response was very slow and displayed persistent photocurrent which the authors reported for a rewritable optoelectronic switch or memory.[162] Similar device structures have been afterwards reported with CVD grown graphene and MoS$_2$, instead of the exfoliated ones.[163] Very recently, Mehew *et al.* reported a graphene-WS$_2$ heterostructure-based photodetector encapsulated in an ionic polymer where WS$_2$ underneath graphene acts as the photoactive layer and the ionic polymer serves as the top gate (Figure 5g).[164] Similar to Graphene-MoS$_2$ devices, they also observed significant photo-gain of $4.8 \times 10^6$, high responsivity of $10^6$ A W$^{-1}$ (Figure 5h) and high detectivity $D^*$ of $3.8 \times 10^{11}$ Jones at a bandwidth of 150 Hz. The key aspect of this work is the significant screening of charge impurities due to the highly mobile ions of the top ionic polymer, which leads to the sub-millisecond response times and a -3 dB bandwidth of 1.5 kHz without the need of gate voltage pulsing.

***Graphene with tunnelling barrier:*** Besides the sensitization of graphene with 2D TMDs, CQDs and perovskite as photo-sensing layers, the introduction of a thin tunnel barrier in graphene double-layer heterostructures can also induce significant photogating effect.[165] The phototransistors comprise a pair of stacked graphene layers (top gate layer; bottom channel layer) sandwiching a thin tunnel barrier (5nm Ta$_2$O$_5$) (Figure 5i). The photo-excited hot carriers can tunnel efficiently into the nearby graphene layer, thereby minimizing hot carrier recombination. The asymmetric tunnelling barrier favours hot electron tunnelling from the top to the bottom graphene layer (Figure 5j). As a result, positive charges accumulate in the top graphene layer, leading to a photogating effect on the bottom graphene transistor, yielding very high responsivity over an ultra-broad spectral range. The electron accumulation in bottom graphene leads to the shift of Dirac point towards negative gate under light illumination shown in Figure 5k. The responsivity can reach 1000 A W$^{-1}$ in visible and 1-4 A W$^{-1}$ from NIR (1.3 µm) to MWIR range (3.2 µm) with bandwidth of 10-1000 Hz, rivalling state-of-the-art mid-infrared detectors and without the need of cryogenic cooling. They noted that the utilization of 2D TMDs (MoS$_2$ or WS$_2$) as tunnelling barrier could further enhance the interlayer hot carrier tunnelling and photogating effect and thus device performance.[165] Zhang *et al*. also reported high responsivity of 8.6 A W$^{-1}$ compared to pure graphene phototransistors from the visible (532 nm) up to the mid-infrared (~10 µm) in a single pure graphene photodetector, by introducing electron trapping centres and by creating a bandgap in graphene through band structure engineering.[39]

**3.3.2. Photogating effect in 2D TMDs based photodetectors**

Despite the very high performance reported in hybrid graphene based phototransistors, their power consumption, electronic read-out schemes and noise are all determined by the absence of bandgap in graphene which leads to large dark current flow. To further improve sensitivity, power consumption and read-out integration, 2D TMDs have been instead considered as potential replacement of graphene for transistor channels sensitized either with CQDs or via doping modulation techniques. The use of semiconducting 2D TMDs channels is of particular promise for they enable the operation of the transistor in the depletion mode, offering thus the advantage of low leakage current in dark conditions by applying appropriate gate voltage.

### 3.3.2.1. Photogating effect in neat 2D TMDs detectors

2D TMDs with a band gap of 1-2 eV have emerged as a promising candidate for next generation of logic transistors, photodetectors and photo-harvesting devices.[62,63] Phototransistors based on a diversity of 2D TMDs including $MoS_2$,[31,166-169] $MoSe_2$,[170] $WS_2$,[171,172] $WSe_2$,[173,174] *etc*. with monolayer or multilayer structures have been reported with a range of performance reported depending on the quality and the nature of 2D-materials investigated. Among them, $MoS_2$ is one of most studied 2D TMDs because of its outstanding properties in view of its high photon absorption efficiency, high carrier mobility up to 200 $cm^2$ $V^{-1}$ $s^{-1}$ and large electrical on/off ratio over $10^8$ as well as natural occurrence of $MoS_2$ single crystals.[30,31] The reported $MoS_2$ phototransistors have shown a very large variation in performance parameters because of the large surface-to-volume ratio, its sensitivity to the surrounding environment and the different substrate treatments. Yin *et al.* demonstrated the first monolayer $MoS_2$ phototransistors with responsivity of 7.5 mA $W^{-1}$ and fast response of 50 ms, the low responsivity is likely due to the absence of any gain mechanism.[166] To further enhance photo-response, many strategies have then been explored employing the photogating effect.

Lopez-Sanchez *et al.* designed top $HfO_2$ gated single layer $MoS_2$ phototransistors with improved contact quality and $MoS_2$ mobility, achieving high responsivity of 880 A $W^{-1}$ but slow response of 0.6-9 s.[31] The high responsivity was attributed to the large photo-gain induced by the trap states in $MoS_2$ or $MoS_2/SiO_2$ bottom interface. Then Furchi *et al*. investigated the origin of photo-response in $MoS_2$, considering both photoconducting and photogating effects.[175] The former contributes to the fast yet low response signal, and the latter is associated with the slow and high-gain response. The long-lived hole traps lying in $MoS_2$ trap states or surface adsorbates (**Figure 6**a) can cause the photogating effect where holes are captured for a certain time while electrons recirculate many times leading to the high photoconductive gain and thus high responsivity. To achieve fast response speed together with high sensitivity, Kufer *et al*. proposed a robust passivation scheme by encapsulating $MoS_2$ with $HfO_2$ or $Al_2O_3$.[50] The isolation from ambient air improved electronic properties with suppressed hysteresis and fast response speed (Figure 6b) and had led to a gate tunable responsivity of $10-10^4$ A $W^{-1}$ and an experimentally measured sensitivity of $10^{11}-10^{12}$ Jones with decay times of 10 ms.

Another efficient strategy for the performance improvement is via gate dielectric engineering. Wang *et al*. integrated multilayer $MoS_2$ phototransistors with ferroelectric P (VDF-TrFE) gate dielectric which favors ultrahigh local electrostatic field in the channel and passivation of the surface trap states. As a result the performance was significantly improved with responsivity of 2570 A $W^{-1}$, fast response of ~2 ms and high sensitivity of ~$10^{12}$ Jones.[176] Interestingly, the spectrum was extended from visible to 1550 nm likely due to the electrostatic field induced defect formation.[176] Lee *et al*. used a semi-transparent and conducting $NiO_x$ as gate electrodes directly on $MoS_2$ channel without an insulator between them (Figure 6c);[177] this device, called metal-semiconductor field effect transistor (MESFET), exhibited extremely high intrinsic-like mobility of 500-1200 $cm^2$ $V^{-1}$ $s^{-1}$ attributed to the reduced charge scattering effect (Figure 6d). They noted that the conducting carriers located at the insulator/$MoS_2$ interface in standard $MoS_2$-on-insulator structures, were unavoidably interfered by the interface traps and gate voltage, while the utilization of $NiO_x$ gate electrodes allowed scattering-suppressed transport. Due to the high mobility, the responsivity has reached as high as 5000 A $W^{-1}$ with fast response time of 2 ms (inset of Figure 6d). The high responsivity with thousands A $W^{-1}$ together with fast speed operation in the above developed phototransistors has been largely due to improved carrier mobility. In addition to TMDs, other types of 2D semiconductors such as group IIIA, IVA, IVB and ternary metal chalcogenides and their photodetectors have also been developed and reviewedextensively elsewhere.[178]

**3.3.2.2. Photogating effect in 2D TMDs sensitized detectors**

Building upon the promising findings of graphene-sensitized phototransistors and in an effort to expand the spectral coverage of 2D TMDs, leveraging at the same time the benefits of those that arise from the presence of a bandgap of around 1-2 eV, in using them as transistor channels, significant efforts were invested in developing high sensitivity infrared photodetectors based on 2D TMDs, sensitized with lower bandgap absorbers.

*MoS$_2$-PbS CQD Hybrid detectors:* In this first report of this class of detectors, a few-layer MoS$_2$ channel was covered by PbS QDs on top (**Figure 7**a). A built-in field was created at the interface between P-type of PbS and N-type of MoS$_2$, which can facilitate the separation of photo-generated carriers in the PbS QD absorber (Figure 7b).[179] The photoexcited electrons transfer into MoS$_2$ and holes remain trapped in QDs, leading to a large gain of $10^6$, high responsivity of $10^5$-$10^6$ A W$^{-1}$ and D* of $5 \times 10^{11}$ Jones with extended spectral sensitivity up to 1.5 µm, determined by the absorption of the QDs. While this work had demonstrated the proof of concept and the fact that an efficient charge separating heterojunction can be formed between QDs and 2D semiconductors, the reported detector was suffering by high dark leakage current because of the serious electron doping of the MoS$_2$ channel during the QD deposition process.[179]

Kufer *et al.* then introduced a thin TiO$_2$ buffer layer between MoS$_2$ and PbS QDs[180] that played a twofold role: it first isolated the MoS$_2$ channel from the environment and second, acted as an efficient photogenerated-electron acceptor buffer layer that funnelled the electrons from PbS QDs to the MoS$_2$ channel. Interestingly, the buffer layer preserves the gate modulation of current in MoS$_2$ by suppressing the high density of localized states at the interface. The on-off ratio has been close to that in pristine MoS$_2$, allowing a dark current as low as few tens of pA (Figure 7c). The device also exhibited fast response of ~10 ms, high EQE of 28% and a record D* of $5 \times 10^{12}$ Jones from visible to NIR spectrum range. The interface engineering presented in this work discloses a new path to control interfaces and doping effects of 2D crystals-based hybrid devices. A similar detector concept based on MoS$_2$ hybrids but with CH$_3$NH$_3$PbI$_3$ perovskite[181,182] or graphene QDs[183] as sensitizers were also reported with improved performance characteristics.

*MoS$_2$-HgTe CQD hybrid detectors:* HgSe and HgTe colloidal quantum dots with even smaller bandgap than PbS, have been considered as a promising low-cost route for mid-IR and far-IR detection due to their tunable bandgap throughout the full infrared spectrum with favourable optical properties.[184-186] The responsivity in their neat photodetectors, however, has been limited to tens of mA/W due to the low carrier mobility, lack of sensitizing centers and the consequent absence of photoconductive gain. Thus, the MoS$_2$-HgTe hybrid phototransistors with TiO$_2$ buffer layer were proposed[187] (Figure 7d). Similar to the MoS$_2$-PbS hybrid case, high gain and responsivity of ~$10^6$ A W$^{-1}$ with fast speed on the order of ms, were achieved benefiting from the synergism of 2D MoS$_2$ and 0D HgTe QDs. By operating in the depletion regime, the noise current was significantly suppressed leading to an experimentally measured D* of ~$10^{12}$ Jones at a wavelength of 2 μm and room temperature (Figure 7e). The spectrum can be further extended by increasing the size of HgTe QDs. The sensitivity was two orders of magnitude higher than prior reports from HgTe-based photodetectors as well as existing commercially available technologies based in extended-InGaAs, InAs or HgCdTe that also require thermo-electric cooling,[9] demonstrating the great potential of hybrid 2D/QDs detector technology in mid-IR applications with compelling sensitivity.

*n-MoS$_2$-p-MoS$_2$ sensitized photodetectors:* MoS$_2$ doping technologies have enabled the polarity transition to P-type and fabrication of MoS$_2$ based PN photodiodes but with.[119-121] Kang *et al.* utilized self-assembled monolayer (SAM)-based doping techniques on 2D TMDs (MoS$_2$ and WSe$_2$). Upon such treatments carrier mobility and photodetection performance were improved by more than one order of magnitude over the pristine control devices.[146] Very recently, a novel device architecture composed of all-2D MoS$_2$ was proposed to serve as an out-of-plane charge separating p-n MoS$_2$ homojunction and an in-plane MoS$_2$ phototransistor.[188] In this case the MoS$_2$ homojunction plays the role of the sensitizing layer for the MoS$_2$ transistor channel. Using low concentrated AuCl$_3$ solution, the top layers of multilayer MoS$_2$ (7-11 nm) were P-type doped while bottom layers remained N-type, thus an out-of-plane PN homojunction was formed (Figure 7f). The out-of-plane PN junction can serve as a sensitizing scheme, which can separate the photo-excited carriers efficiently and produce a photo-gain of >10$^5$ electrons per photon. The high gain and fast response of ms level taken together with low noise yields record and gate-tunable sensitivity up to $3.5 \times 10^{14}$ Jones at bandwidth of 1 Hz and 10 Hz (Figure 7g). The same device concept can be in principle applied in other 2D semiconductors, particularly those of low bandgap, such as BP, that hold promise to extend the spectral coverage of the 2D materials realm.

### 3.3.3. 2D photodetectors sensitized by plasmonic nanostructures

An alternative to semiconductor sensitizers for 2D phototransistors has been based on plasmonic metal nanostructures. Very intense and resonant absorption enhancement can be achieved by utilizing the strong local field concentration from plasmons at the resonant wavelength of the nanostructures. Upon decoration with metallic nanoparticles or nanoantenna arrays, the photocurrent in silicon and 2D materials-based photodetectors has been reported to increase at the plasmonic resonance of the sensitizers, enabling narrowband spectrally selective photodetectors.[189-192] For example, in a few-layer MoS$_2$ detector sensitized with Au nanostructures, the recorded photocurrent was improved due to absorption enhancement in the MoS$_2$ from the near field of Au nanoparticles.[193-196] Graphene nanodisks and Au plasmonic nanoantennas have also been reported to enhance the absorption efficiency in graphene from less than 3% to 30%.[197-202]

Another important property of plasmonic sensitizers is their ability to generate energetic or "hot" carriers, which can enable photocurrent generation from photons with energy below the bandgap of the 2D semiconducting channel. The incident light can couple into surface plasmons by nanoantennas and the nonradiative decay of the plasmons results in hot electrons that can transfer across the Schottky barrier at the metal–semiconductor interface and be detected as a photocurrent[203-206] (**Figure 8**a). Hot-electron-induced photodetection has been reported in 2D graphene[200,201] and $MoS_2$.[150,207,208] By scanning the laser beam along the $MoS_2$ channel (Figure 8b), the spatially resolved scanning photocurrent with different photon energies was recorded as shown in Figure 8c. The strongest photocurrent response was observed at the $MoS_2$-metal junction with extended spectrum coverage up to 1.55 µm due to the generated hot electrons from the metal to $MoS_2$.[207] A bilayer $MoS_2$ integrated with a plasmonic antenna array also exhibited sub bandgap photocurrent with a photo-gain of $10^5$ and responsivity of 5.2 A W$^{-1}$ at 1070 nm, the gain in this case has been attributed to charge trapping in $MoS_2$/metal or $MoS_2$/$SiO_2$ interfaces.[150] Based on the same concept a graphene-antenna sandwich photodetectors (Figure 8d) showed 800% enhancement of the photocurrent due to the hot electron transfer and direct plasmon-enhanced excitation of intrinsic graphene electrons.[200] When measuring the local photocurrent in the device, by performing a line scan of the excitation laser between the source and drain electrodes, an antisymmetric photocurrent response is observed (Figure 8e), showing that the plasmonic antenna particularly heptamer structures provide larger field enhancements, multiple hot spots and a greater yield of hot electrons. While intriguing, the concept of having resonant selective photodetectors based on plasmonic sensitizers has, till now, led to much lower performance over the semiconductor-sensitized counterparts due to the challenge of competing efficiently over the ultrafast hot carrier relaxation in metals and the generated heat losses.

**3.3.4. 2D-organic hybrid based photodetectors**

Organic semiconductors possess some very favorable properties particularly relevant in wearable (opto)electronic applications, in view of their optical properties, stretchable/flexible form factor and production-scalable methods.[209-211] An opportunity therefore exists in synergistically combining those with 2D materials for optoelectronic applications. Particular benefits arise from the compatibility of those two material platforms: 2D atomic crystals, for example, provide atomically flat and inert surfaces, that can be ideal for ordered self-assembly of organic molecules.[212] Moreover, high quality organic layers can form on top of 2D materials via thermal evaporation or spin/dip-coating in view of the vdW force interactions and the absence of dangling bond formation at their interface. The vdW interactions can allow the epitaxial growth of organic-based films with larger crystal grain size, thus enhancing the electronic properties of organic semiconductors. Graphene and h-BN have been demonstrated as an ideal template or dielectric substrate for $C_{60}$ film, dioctylbenzothienobenzothiophene (C8-BTBT) and rubrene transistors with high carrier mobility exceeding 10 $cm^2$/Vs.[213-216] Due to the efficient charge transfer at interface, the organic molecules can significantly tune the Dirac point of graphene, and change the transport behaviour from p-type to ambipolar and finally n-type, offering new functional devices.[217] Ultrafast charge transfer (6.7 ps) and long-lived charge-separated states (5.1 ns) have been observed in pentacene-$MoS_2$ p-n heterojunctions.[218] By employing plasmonic metasurfaces in hybrid $MoS_2$-organic heterojunction, the charge generation within the polymer is enhanced 6-fold and the total active layer absorption band is increased.[219] These features suggest significant promise for 2D-organic hybrid heterostructures in photovoltaics and photodetectors. Recently, a new graphene-organic semiconductor based vertical field effect transistors (VFETs) (**Figure 9a**) was demonstrated to exhibit high on-off ratio up to $10^5$ (Figure 9b) thanks to the partially-screened field effect and selective carrier injection through graphene.[220,221] Solution-based graphene nanomaterials were modified with organic molecules to open a bandgap due to the charge redistribution between the C-C bonds, which in turn has led to promising photodetector performance characteristics and high responsivity of 2.5 A/W in the mid-infrared spectral region (3-5 $\mu$m).[222] Similar to 2D/QDs detectors, the organic molecules or polymers can also be used to sensitize the surface of 2D channels thus producing a significant photoconductive gain through the photo-gating effect. Lee *et al.* fabricated a hybrid photodetector comprising organic dye molecules (rhodamine 6G) and graphene with broad spectral photo-response and responsivity of 460 A W$^{-1}$.[223] Liu *et al.* have epitaxially grown small molecule C$_8$-BTBT on top of graphene using a CVD approach and fabricated C$_8$-BTBT/graphene hybrid phototransistors which can exhibit photo responsivity of 10$^4$ A/W, photo-gain larger than 10$^8$ and time response of 25 ms.[224] Under the same device concept, graphene-polymer semiconductor (P3HT or PTB7) hybrid phototransistors (Figure 9c) have exhibited responsivity exceeding 10$^4$ A W$^{-1}$ and fast temporal response of ~7.8 ms. In this device, the use of a self-assembled-monolayer (SAM) functionalization to effectively remove surface traps and charged impurities between graphene and SiO$_2$ substrate, further improved responsivity up to 10$^5$ A W$^{-1}$ (Figure 9d).[225,226] Tan *et al.* demonstrated that the utilization of piezoelectric (PZT) substrate can improve the photocurrent of graphene-P3HT photodetectors by 10 times compared to that based on SiO$_2$ substrate due to the enhanced separation of photogenerated electrons and holes under the electric field of the polarization from the piezoelectric substrate.[227] It is noted that carbon nanotubes were also combined with graphene to form hybrid photodetectors with broadband spectal response (covering 400–1,550 nm), high responsivity of 4100 A W$^{-1}$ and a fast response time of 100 $\mu$s.[228,229] Besides graphene, 2D TMDs such as MoS$_2$ and WS$_2$ have also been combined with organic materials to achieve hybrid photodetectors.[230-232] For instance, Yu *et al.* reported dye-sensitized MoS$_2$ photodetectors utilizing a single-layer MoS$_2$ treated with rhodamine 6G (R6G) organic dye molecules. The hybrid devices exhibited responsivity of 1.17 A W$^{-1}$, detectivity of 1.5×10$^7$ Jones, and spectral converage up to 980 nm.[233] Overall, 2D-organic semiconductor hybrid systems have attracted a lot of attention and the recent progress in the field has been extensively described elsewhere.[234,235]

## 4. Demonstrated applications of 2D based photodetectors

## 4.1. Flexible electronics and detectors

Flexible electronics and sensors have opened a new realm of functionalities in bendable and portable device technologies and applications such as wearable health monitors, electronic skins, portable touch panels as well as Internet-of-things sensors and transducers seamlessly integrated in an ubiquitous manner. Conventional semiconductors (a-Si) and metal oxide thin-films have been explored and currently used in flexible electronics, yet some of their limitations such as opaqueness, thickness scalability and high-cost manufacturing[236,237] need to be overcome for high performance, low-power consumption, small footprint devices. The nanoribbon or nanomembranes based on single crystalline inorganic materials have emerged as main contenders in high performance bendable and stretchable electronic applications.[238] Organic based semiconductors have also been recently developed particularly in transparent and stretchable optoelectronics for biomedical applications such as electronic skins, health monitoring, medical implants and human-machine interface.[209-211,239-242] Efforts on more robust and higher carrier mobility of organic electronics are still ongoing.[243,244] Along this direction, 2D materials offer an alternative promising route towards these goals.[21,245] Leveraging its extraordinary optical and electronic properties and its large-scale manufacturability, graphene has been widely reported as soft and transparent electrode in modern flexible electronics.[246,247] But the absence of bandgap limits the utilization of graphene in digital electronics. To address this, graphene-organic semiconductor based VFETs were demonstrated to exhibit high-off ratio up to $10^5$ with a bending radius of < 1mm.[220,221] Meanwhile, 2D TMDs such as $MoS_2$ have been proposed and developed in logic circuit components for flexible electronics.[248-250] The $MoS_2$ transistors on flexible substrates have shown ON/OFF ratios over $10^7$ and mobility of ~30 cm$^2$ V$^{-1}$ s$^{-1}$, which is on par to that on rigid substrate.[248] 2D heterostructures combined with graphene as electrodes, TMDs as channel materials and hBN as gate dielectric have also been demonstrated as promising architectures for all-2D based flexible electronics and detectors (**Figure 10**a).[249,251,252] The graphene/MoS$_2$ flexible photodetectors have reached responsivity of 45.5 A W$^{-1}$ with a gain of ~10$^5$ even upon bending curvature of 1.4 cm.[253] Liu *et al.* fabricated fully transparent and flexible graphene P-N junction-based IR photodetectors through a facile chemical doping technique.[254] Recently, highly stretchable graphene devices by intercalating graphene scrolls in between graphene layers (Figure 10b) were also reported, where the graphene scrolls provide conductive paths to bridge cracks in the graphene sheets, thus maintaining high conductivity under strain. The corresponding all-carbon based flexible transistors exhibited a transmittance of >90% and retained 60% of their original current output at 120% strain, showing superior performance in terms of mobility, on/off ratio and being highly stretchable.[255] BP-based transistors on highly bendable polyimide substrate (Figure 10c) were reported with mobility up to 310 cm$^2$ V$^{-1}$ s$^{-1}$ and excellent mechanical durability.[256,257] Hybrid flexible photodetectors based on WSe$_2$, graphene-carbon nanotube and ZnS-MoS$_2$ heterojunctions were also reported with responsivity of 50 A W$^{-1}$ and fast response (~40 ms).[258-261] 2D materials are expected to play a big role in the next generation of flexible electronics and optoelectronics in view of their atomically thin form factor and their outperforming electronic properties over those of current flexible electronic semiconductors.[21,22,262]

### 4.2. Silicon and CMOS Integration

Conventional silicon semiconductor technology has been the cornerstone of modern electronics and may continue so for the near future. One of the first generation of applications of 2D materials may therefore be in combination with standard silicon technology in providing new functionalities to silicon or replacing some manufacturing processing steps needed to implement junctions on silicon. 2D materials combined with conventional Si has therefore been explored and demonstrated a series of promising electronic and photonic devices. Graphene barristors were realized by combining silicon to form a graphene-silicon Schottky barrier.[263]

Large modulation on the barristors current with on/off ratio of $10^5$ and ideality factor of 1.1 was achieved by adjusting the gate voltage to control the barrier height, which overcame the key obstacle in graphene-based electronics. Recently, Wan *et al.* presented a self-powered, high-performance graphene-enhanced ultraviolet silicon Schottky photodetector with a $Al_2O_3$ anti-reflection layer (Figure 10d). At zero-biasing (self-powered) mode, the photodetectors exhibit high photo-responsivity (0.2 A W$^{-1}$) in the ultraviolet region, fast time response (5 ns) and high specific detectivity ($1.6 \times 10^{13}$ Jones) comparable to that of state-of-the-art Si, GaN, SiC Schottky photodetectors. They also demonstrated high stability of the device exceeding 2 years.[264] $MoS_2$ has also been integrated with silicon to form a heterojunction that has then been exploited in silicon photodiodes, eliminating the need of doping silicon.[265-268] $MoS_2$-Si heterojunctions exhibit high sensitivity of $10^{13}$ Jones with fast response times of ~3 μs due to the efficient built-in field at their interface.[266] Lopez-Sanchez *et al*. reported an avalanche photodiode based on $MoS_2$/Si heterojunctions with multiplication factor exceeding 1000 and respectably low noise floor, enabling low-noise photon counters in 2D APDs.[46] The integration of graphene detectors with a silicon-on-insulator waveguide (Figure 10e) gave rise to the enhancement of graphene absorption and thus photodetection efficiency through coupling the evanescent field from the optical waveguide mode to the graphene absorber, preserving both high speed and broad spectral bandwidth.[269] These detectors had responsivity exceeding 0.1 A W$^{-1}$ with response rates exceeding 20 GHz and were demonstrated successfully in an instrumentation-limited 12 Gbit/s optical data link. Similarly, Youngblood, *et al.* also reported a multilayer black phosphorus photodetector integrated on a silicon photonic waveguide. High responsivity of ~0.6 A W$^{-1}$ with large bandwidth over 3 GHz operating in the near-infrared telecom band was achieved.[41] A silicon waveguide-integrated bilayer $MoTe_2$ was recently demonstrated to act as both a light source and a photodetector on a single chip platform (Figure 10f),[270] paving the way for Si-photonics enabled by 2D materials. Besides silicon, the integration of $MoS_2$ on germanium (Ge) was reported in a band-to-band tunnelling transistor (Figure 10g) with subthreshold swing as low as 3.9 mV/decade and an average value of 31.1 mV per decade which overcome the theoretically predicted limits in conventional bulk transistor configurations.[271]

Complementary metal–oxide semiconductor (CMOS) based integrated circuits are at the heart of the technological revolution in modern society, CMOS integration of any new technology introduced warrants its easier industrial acceptance and market entry leveraging the mass production capabilities and low cost manufacturing processes developed for CMOS electronics. Individual CMOS-compatible graphene photodetectors covering the fibre-optic telecommunication bands[272] and graphene/silicon CMOS hybrid hall integrated circuits (ICs) via low temperature process[273] have been reported. Very recently, graphene-QD hybrid photodetectors have been integrated with CMOS read-out circuits to demonstrate a broadband CMOS-based digital camera capable of capturing both visible and infrared light.[274] The schematic diagram of CMOS integration of graphene with 388 × 288 pixel image sensor read-out circuit is shown in **Figure 11**a. Graphene was transferred on to a CMOS die and connected with the bottom readout circuity by vertical metal interconnects. Then graphene was patterned to define the pixel shape followed by sensitization of PbS QDs on top (Figure 11b). This hybrid technology has reached high responsivity of $10^7$ A W$^{-1}$, measured D* of $10^{12}$ Jones and fast switching speed of 0.1-1ms that were achieved in the spectral range of 300-1800 nm. The successful integration of 120,000 pixel photodetectors in a single focal plane array has enabled high-resolution imaging in a broadband spectrum under weak light atmospheric conditions or environmental conditions that standard silicon cameras would have not been able to capture. (Figure 11c). Zanjani *et al.* also demonstrated monolithically integrated CMOS-graphene gas sensors combining the superior gas sensitivity of graphene with the low power consumption and low cost silicon CMOS platform.[275] The graphene–CMOS integration is pivotal for incorporating 2D materials into the next generation of sensor arrays and CMOS imaging systems. The successful integration of 2D materials on a silicon photonics and microelectronic platform with high performance,[276] offers the possibility of CMOS integrated graphene and related 2D materials (TMDs, BP, etc.) for commercial optoelectronic applications.

## 5. Future outlook

In this concluding section, we first compare the performance between typical 2D and hybrid based photodetectors and commercial silicon, germanium, InGaAs and HgCdTe photodetectors.[9] **Figure 12a** shows the responsivity versus the response time, demonstrating comparable performance. Graphene and BP photodetectors exhibit ultrafast speed and comparable responsivity. For hybrid and TMDs based photodetectors possessing gain, the responsivity is much higher but the response speed is limited on the order of milliseconds. Figure 12b also shows the detectivity versus the wavelength for different classes of photodetectors. The detectivity of 2D based photodetectors is comparable or even superior than that of conventional ones in the SWIR spectral range at room temperature. Overall, there is still a large room for improvement for 2D based photodetectors particularly aiming at faster response speed and extension of their spectral range.

In the following, we discuss in brief the future outlook and some of the challenges needed to be addressed to further boost performance, technology and manufacturing readiness levels in order to promote 2D material photodetector technologies towards commercial applications.

### 5.1. Improvement in linear dynamic range

The linear dynamic range (*LDR*) of graphene photodetectors are limited to only 7.5 dB due to their intrinsic hot-carrier dynamics, which causes deviation from a linear photoresponse at low incident powers.[29] TMDs and TMD/QDs hybrid based photodetectors also suffer from relatively low LDR due to their dependence of dynamic range on the density of the sensitizing states as well as the parabolic nature of the density of states in the TMDs.[31,180,187] As compared to 2D materials, state-of-the-art conventional (Si, Ge, GaAs, *etc.*) photodetectors currently exhibit a linear response over a larger range of optical powers. High *LDR* in 2D-based photodetectors is thus needed for high-resolution sensing and video imaging applications. To address this issue in graphene detectors, Sanctis *et al.* defined p-p´ junctions by laser-assisted displacement of $FeCl_3$ in $FeCl_3$ intercalated few-layer graphene, these junctions can quench the hot-carrier effects and exhibit photocurrent signals from purely photovoltaic (PV) effect. As a result, they achieved an extraordinary LDR of 44 dB which is 4500 times larger than that of previously reported graphene photodetectors.[277] Their detectors also exhibited high stability under atmospheric conditions and a broad spectral response from ultraviolet to mid-IR wavelength. In 2D-QDs hybrid phototransistors, Nikitskiy *et al.* employed an electrically active QDs photodiode as the sensitizing element instead of a passive sensitizer atop graphene and achieved a significant enhancement of the *LDR* up to 110 dB, which was evidenced by a flat responsivity across the wide range of power density.[154]

**5.2. Contact and mobility engineering**

Low contact resistance in 2D material-based electronics and optoelectronics is critical for achieving efficient carrier injection, high responsivity and high frequency operation.[31,278] However, a large Schottky barrier is usually formed between 2D materials and conventional 3D metals due to the lack of chemical bonding on the chemically inert 2D surface, the large energy offset and Fermi level pinning at the interface.[279] Although optimized metal contacts have been studied in 2D materials,[280,281] new contact schemes towards lower contact resistance are needed to further enhance the device performance. Due to the finite density of states and its chemical inertness, graphene has been reported as an ideal electrode on other 2D semiconductors such as $MoS_2$ with a barrier-free contact via tuning the work function of graphene. The low Ohmic contact resistance leads to higher photo-response and measured field effect mobility.[282-284] The one-dimensional edge contact geometry was also developed to make a high-quality electrical contact between graphene and metal, leading to outperforming electronic performance such as the phonon-scattering limited room-temperature mobility up to 140,000 $cm^2$ $V^{-1}$ $s^{-1}$.[285] Recently, low-resistance metal–semiconductor contacts (200-300 Ω µm) were also obtained by interfacing semiconducting 2H-MoS$_2$ layers with 1T-MoS$_2$.[286] The utilization of thin TiO$_2$ layer between MoS$_2$ and metal was also reported to suppress trap state density at the contact interface and thus enhance both extracted mobility and photoresponsivity.[287] Overall, the performance of 2D photodetectors can be drastically improved by these contact engineering schemes due to more efficient photo-carrier extraction and recirculation through the external circuit.

Mobility is another important challenge in 2D materials that determines to a large extent their performance as transistors or photodetectors. The utilization of high-ƙ dielectrics as strongly coupled top-gate insulators has improved the room temperature mobility up to the order of 100 cm$^2$ V$^{-1}$ s$^{-1}$ due to dampening of Coulomb scattering from the charged traps or impurities.[288,289] This taken together with the low contact resistance through contact engineering, as discussed above, is expected to lead to even higher extracted mobility values for 2D-TMD transistors. In addition, strain effects have been theoretically demonstrated to improve the mobility in MoS$_2$ by one order of magnitude through the suppression of electron-phonon coupling.[290] The mobility engineering in 2D materials is very promising in terms of realization high-speed and high-sensitivity 2D photodetectors. For example, higher mobility leads to faster photocarrier transit time from source to drain, enabling higher photoconductive-gain for a given carrier lifetime. Besides mobility, the subthreshold swing and ON/OFF ratio in 2D based transistors particularly 2D/QDs hybrids phototransistors also play a critical role since they determine the ultimate sensitivity and linearity of corresponding photodetectors when operating around the threshold voltage.[49]

### 5.3. Challenges in 2D hybrid platforms

2D hybrid systems employing synergistically other materials such as 2D TMDs, CQDs or perovskites have already shown performance on par with detector technology based on silicon and InGaAs for the Vis−SWIR range and have the potential to outperform them in sensitivity and cost. Graphene-PbS CQD hybrid phototransistors have reached specific detectivity of $7\times10^{13}$ Jones with fast speed below 1ms from visible to SWIR range, enabling CMOS compatible, broadband, high-resolution imaging systems.[274] MoS$_2$ based phototransistors have also exhibited detectivities on the order of $10^{12}$-$10^{14}$ Jones at visible and SWIR beyond 2 µm, realized by chemical doping of MoS$_2$ or HgTe CQD sensitization.[187,188] Despite the great progress in this new photodetector platform, significant further improvements are to be expected in terms of optimizing 2D transistor channels, sensitizers and interface towards more sensitive, broadband and faster photodetectors.

In 2D hybrid platforms, the 2D channels act as carrier transport layers and therefore must meet the high mobility and gate current modulation features, which contribute to the high gain and low noise, respectively. Graphene has very high mobility but being a semimetal also suffers from large dark current. Strategies to open a bandgap in graphene such via doping or nanostructuring can give rise to lower noise and higher sensitivity. 2D TMDs on the other hand, albeit limited by lower carrier mobility, they benefit from very low dark current when operated in depletion mode, offering the potential for overall improvement in the sensitivity of the detector. Potential pathways to improve their performance lie on enhancing their mobility by low contact resistance or mobility engineering using high-ƙ dielectrics. Alternative or synergistically to that, the use of new emerging semiconducting 2D materials such as BP, with a moderate bandgap of 0.3 eV in bulk and high mobility of 1000 cm$^2$ V$^{-1}$ s$^{-1}$, may also be considered.[36] 2D topological insulators such as Bi$_2$Se$_3$ with symmetry protected surface states and insulating or semiconducting internal bulk band structure have been reported with high mobility up to 1750 cm$^2$ V$^{-1}$ s$^{-1}$ at room temperature, posing them as new promising material platforms for transistor channels towards broadband, highly sensitive photodetectors.[145,291]

As far as the sensitizers is concerned, a high absorption coefficient with large and ideally tunable spectral coverage is a prerequisite. In addition to this and in order to facilitate efficient charge transfer to the transistor channel, thereby high quantum efficiency, the sensitizer layer should also possess favourable electronic properties in terms of carrier diffusion length, carrier mobility, doping and minority carrier lifetime. Colloidal quantum dots in that aspect, can serve as strong absorbers with broadband spectral range from the visible to the infrared region. Particularly the HgSe and HgTe QDs can cover the full spectrum up to 20 µm,[184-186] which offers the possibility of 2D/QDs hybrid systems with sensitivity extending in the MWIR and even LWIR. The type-II band alignment between the CQDs sensitizer layer and the 2D channel can also be engineered by using different capping ligands, which can further extend the depletion width and improve the photocarrier transfer toward the transport channel thereby enhancing the charge collection efficiency.[292] Recently, solution-processed perovskites were also demonstrated with strong light absorption and remarkable diffusion lengths exceeding 10 µm,[293] suggesting them an ideal sensitizer to reach high optical absorption and high quantum efficiency, although limited in the visible and near infrared part of spectrum. In the future, new environmentally friendly CQDs with smaller bandgap, such as $Ag_2Se$ QDs with absorption wavelength up to 6.5 µm[294] or other semiconducting 2D materials such as BP can be employed as sensitizers in the 2D hybrid based IR detectors.

The interface between 2D channels and sensitizers plays a key role for the performance of the hybrid detectors. The introduction of a thin $TiO_2$ buffer layer has been instrumental in avoiding the electrical doping on the channel from the environment and allowing nearly full modulation of its conductivity upon sensitization.[180] The use of self-assembled monolayers at the interface may also lead to efficient cross-linking between QDs and 2D channels and simultaneously passivate electronic defect states.[295,296] The transformation of the sensitizer layer from an electrically passive one to an active one can really lead to new configurations and even better performance. This concept has been demonstrated in a graphene – PbS QD photodetector in which the PbS layer, by the use of a top electrode has been transformed into a Schottky photodiode.[154] As a result the charge transfer efficiency from the sensitizer to the graphene was improved leading to an EQE of 80% and the linear dynamic range of the detector increased to 110 dB. Such an approach led to a detector that provided the combined benefits of a photodiode and a phototransistor. Based upon this concept several new photodetector architectures can be implemented as a sensitizing layer, such as p-n or heterojunction diodes leading to even better characteristics in terms of speed, noise *etc*. This concept can also be applied in 2D TMDs-QDs hybrid systems.

**5.4. Noise suppression**

The suppression of noise in photodetectors is a very demanding and challenging task yet it is required for high sensitivity. Generally, electrical noise is composed of four intrinsic sources: thermal or Johnson noise, shot noise, generation-recombination (G-R) noise and flicker or 1/f noise. Thermal and shot noise components are frequency independent (thus typically called white noise) and originate from the random motion of charge carriers. The spectral density of thermal noise is given by Nyquist's formula $S_I(f) = 4k_BT/R$, where $k_B$ is Boltzmann's constant and T is temperature, R is resistance of detector. Whereas the shot noise is determined by Schottky's theorem $S_I(f) = 2q<I>$, where $<I>$ is the average value of the electrical current. 1/f noise is frequency dependent and caused by electrostatic fluctuations resulting in charge carrier number fluctuation or mobility fluctuation or their combination. The origin of 1/f noise is not always known but has been ascribed, depending on the material platform, to metal-semiconductor interfaces, trap states or effects from semiconductor edges and dislocations.[297] G-R noise bulges are superimposed on the 1/*f* spectrum at low frequency and associated with trap states of different time constants.

In 2D material-based photodetectors, the trap states in the substrate or the gate oxide can capture or emit carriers from/to atomically thin channels leading to large current fluctuations and significant 1/*f* noise which currently determine the sensitivity of those detectors.[297] Bilayer or few layer graphene has shown lower 1/*f* noise than single layer due to their stronger electrostatic screening ability,[298] whereas suspended graphene has exhibited very low noise due to the removal of the oxide substrate effects underneath the channel.[299] Defect-free and atomically flat 2D h-BN as a substrate has also been reported to lead to noise suppression in graphene by 10-100 times over neat graphene devices.[300] Further improvements can be reached by encapsulating graphene with h-BN on both sides and by making one-dimensional edge contacts[285,301] In semiconducting 2D TMDs, ultralow dark current can be achieved in view of the atomically thin profile of the channel and by operating the transistor in depletion mode, thus shot noise is not expected to be a limiting noise factor in these devices. However, the 1/f noise is still significant due to the trap states, environmentally induced Coulomb scattering centers as well as contact barriers.[302,303] The strategies for suppressing noise in graphene can also be applied in TMD-based devices and this remains to be demonstrated. High quality contact and encapsulation with high-ƙ dielectric were demonstrated to suppress the noise in $MoS_2$.[304,305] Besides, for CVD grown 2D graphene and TMDs, the existing numerous defects and grain boundaries can act as scattering centers that significantly increase the 1/f noise.[306,307] The growth of high quality 2D materials with lower lattice disorder remains a challenge that needs to be addressed also for low noise and high sensitivity detectors. Although 1/*f* noise is useful to detect environmental chemical species,[308] it ultimately limits the sensitivity of detectors. Readers can see recent reviews focused on noise in 2D materials.[297,309]

## 5.5. Wafer-scale production

To realize widely commercial applications of 2D materials in modern electronics and optoelectronics, wafer-scale growth or processing is instrumental. So far, many groups have developed growth methods towards wafer-scale production of 2D materials. As one of the most studied 2D materials, large area graphene with >95% monolayer uniformity and low defect density has been achieved on 300 mm Si wafer covered with one copper layer using chemical vapour deposition (CVD) method,[310,311] and now it is already commercially available. Polycrystalline graphene with 30-inch size (**Figure 13**a) can also be produced on flexible substrates such as polyethylene terephthalate (PET) via roll-to-roll processing (Figure13b),[312] enabling massive high-throughput processing of flexible electronics. 2D TMDs, as semiconducting films with high quality, scalable size and controllable thickness, have also been grown using the CVD method.[79-81] Kang, *et al*. achieved 4-inch wafer-scale films of monolayer $MoS_2$ and $WS_2$ (Figure 13c), grown directly on insulating $SiO_2$ substrates using a newly developed metal–organic chemical (MOCVD) method (Figure 13d).[35] The resultant monolayer films possess excellent spatial homogeneity over the entire wafer and appreciable mobility of 30 $cm^2$ $V^{-1}$ $s^{-1}$ at room temperature. Wafer-scale growth of other 2D semiconducting materials such as $WSe_2$,[313] $InSe$[314] and $GaS$[315] has also been demonstrated pointing to the generalization of wafer scale manufacturability of 2D materials. As an important dielectric, few layer h-BN has also been reported based on wafer-scale and wrinkle-free epitaxial growth using high-temperature and low-pressure CVD method.[316] This demonstrated progress on wafer-scale growth of a variety of 2D materials keeps up the promise for the next generation of low cost, easy-processing, flexible and CMOS compatible optoelectronic platforms enabled by 2D materials. Yet, CVD grown 2D materials still exhibit lower quality than small-scale exfoliated counterparts, thus further improvement is still foreseen in terms of suppressed lattice disorder, higher mobility and stronger photoluminence for CVD grown 2D materials. Efficient transfer techniques have been developed[317-319] to stack 2D building blocks towards multifunctional heterostructures, yet issues related to wrinkle formation, contamination and precise control of stacking orientation during the transfer process should be addressed to achieve such functionalities and exploit the full potential of the 2D family, in large scale processes. An alternative approach to large scale manufacturing may rely on printing from solution, a method that may be useful when cost is critical and the requirements for quality are more relaxed. In this aspect, Coleman *et al*. produced large amount of 2D nanosheets in liquid phase by liquid exfoliation method and dispersed them in common solvents, which can be deposited as individual flakes or formed into films.[320] Carey *et al*. demonstrated fully inkjet-printed 2D-material active heterostructures with graphene and h-BN inks, and fabricated all inkjet-printed flexible and washable field-effect transistors.[82] McManus *et al*. recently proposed a new approach to achieve inkjet-printable, water-based, two-dimensional crystal formulations including graphene, TMDs and h-BN (Figure 13e), and fabricated high performance of logic memory arrays (Figure 13f) based on multi-stack films.[83]

In summary, we have summarized concisely the recent progress in the field of photodetectors based on 2D materials including graphene, TMDs and BP as well as their heterostructures also hybridly with 0-dimensional or 3-dimensional material platforms. The presence of photoconductive gain or not has led to the classification of two classes of 2D-based photodetectors. One is heterojunction based photodiodes with both in-plane and out-of-plane configurations, which operate as photodiodes exhibiting fast response and high quantum efficiency. The photo-thermoelectric and bolometric detectors can also be applied in 2D materials particularly graphene, enabling the broadband detection from MWIR to submillimetre wavelength range. Another is hybrid based phototransistors, possesing photoconductive gain by combining 2D materials or CQDs, perovskite, metal nanostructures and organic semiconductors. The high performance taken together with the advantages of low-cost and easy-processing, flexibility, integrability with silicon technologies or CMOS compatibility, have lend 2D materials a very promising future in next generation of photodetector applications.

**Acknowledgements**

We acknowledge financial support from the European Research Council (ERC) under the European Union's Horizon 2020 research and innovation programme (grant agreement No 725165), the Spanish Ministry of Economy and Competitiveness (MINECO) and the "Fondo Europeo de Desarrollo Regional" (FEDER) through grant TEC2017-88655-R. This work was also supported by by European Union H2020 Programme under grant agreement n°696656 Graphene Flagship. We also acknowledge financial support from Fundacio Privada Cellex, the CERCA Programme and the Spanish Ministry of Economy and Competitiveness, through the "Severo Ochoa" Programme for Centres of Excellence in R&D (SEV-2015-0522).

Received: ((will be filled in by the editorial staff))
Revised: ((will be filled in by the editorial staff))
Published online: ((will be filled in by the editorial staff))

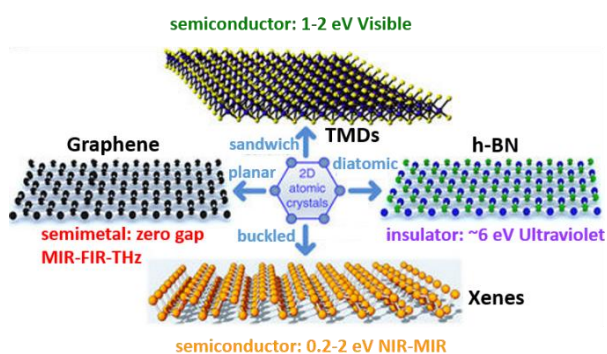

**Figure 1.** Crystalline structures of 2D atomic crystals including graphene, TMDs, h-BN and Xenes. Reproduced with permission.[21] Copyright 2014, Nature Publishing Group.

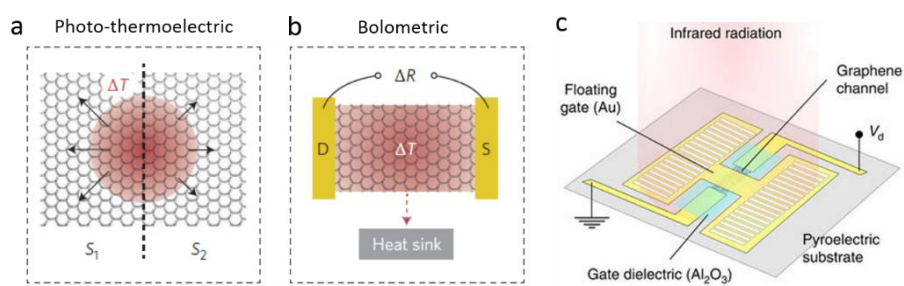

**Figure 2.** Schematic representation of the photo-thermoelectric (a) and bolometric (b) effect in graphene detectors. The red shaded area indicates elevated electron temperature with the temperature gradient ($\Delta T$) and the resistance change across the channel ($\Delta R$); $S_1$ and $S_2$ are the Seebeck coefficient in graphene areas with different doping. (c) Scheme of the graphene pyroelectric bolometer, where the conductance of a single layer graphene channel is modulated by the pyroelectric substrate and by a floating gate. (a,b) Reproduced with permission.[100] Copyright 2014, Nature Publishing Group. (c) Reproduced with permission.[104] Copyright 2017, Nature Publishing Group.

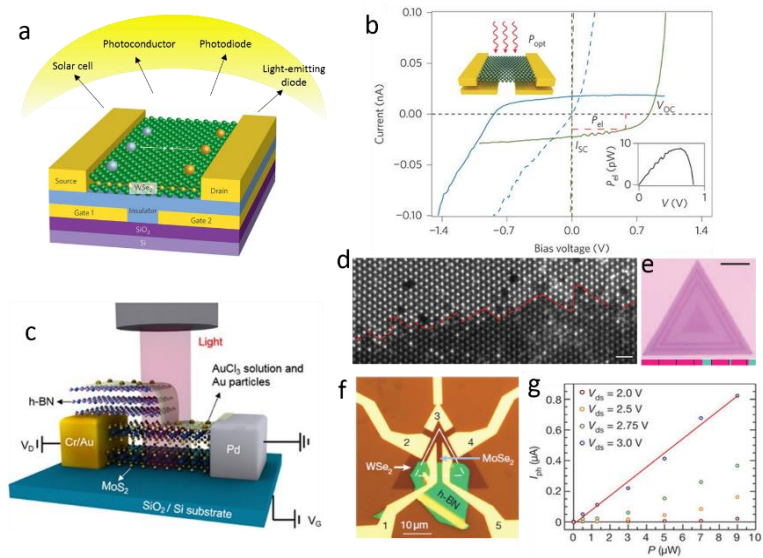

**Figure 3.** (a) Schematic diagram of lateral monolayer WSe$_2$ p-n diode with split-gate electrodes for varying optoelectronic applications. The metal gates underneath are separated by an insulator from WSe$_2$ channel. (b) I–V characteristics of the WSe$_2$ devices under optical illumination with 1,400 W m$^{-2}$. Top inset: Schematic of experiment. Lower inset: electrical power $P_{el}$ versus voltage under incident illumination. (c) Three-dimensional schematic diagrams of lateral MoS$_2$ p-n diodes using chemically doping. (d) Atomic-resolution Z-contrast STEM images of the in-plane interface between WS$_2$ and MoS$_2$ domains. The red dashed lines highlight the atomically sharp interface along the zigzag-edge direction. (e) Optical images of five-junction MoSe$_2$-WSe$_2$ heterostructures. The scale bar is 10 µm. (f) Micrograph of a MoSe$_2$-WSe$_2$ single junction based device. (g) Photocurrent $I_{ph}$ as a function of the illumination power $P$. The red line is a linear fit, indicating a linear dependence of $I_{ph}$ on $P$ at high bias. (a) Reproduced with permission.[115] Copyright 2014, Nature Publishing Group. (b) Reproduced with permission.[112] Copyright 2014, Nature Publishing Group. (c) Reproduced with permission.[122] Copyright 2014, American Chemical Society. (d) Reproduced with permission.[125] Copyright 2014, Nature Publishing Group. (e-f) Reproduced with permission.[127] Copyright 2018, Nature Publishing Group.

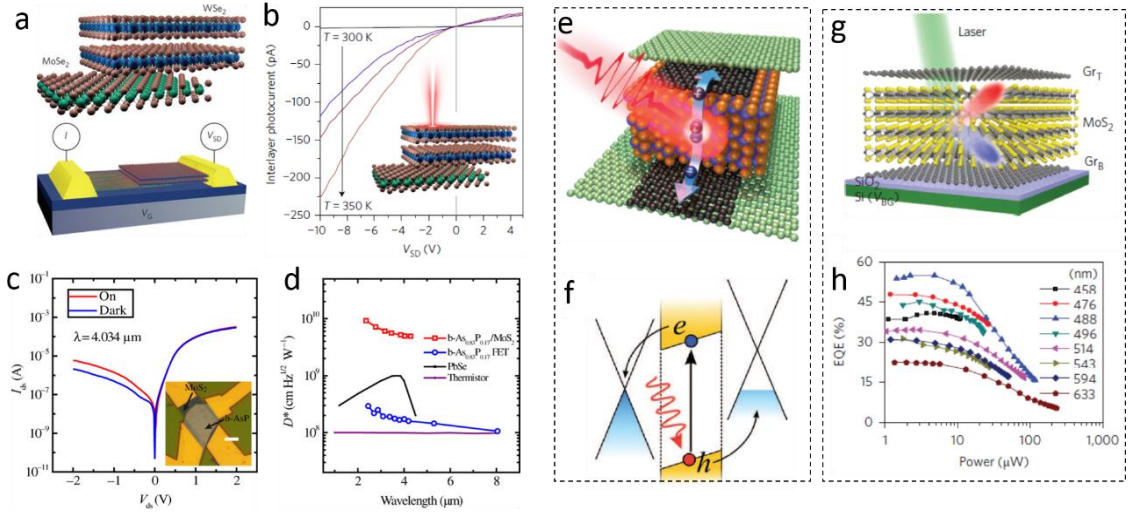

**Figure 4.** (a) Schematics of the atomic-layer heterostructure (top) and the n$^+$-n heterojunction device (bottom). (b) Interlayer photocurrent vs $V_{SD}$ at $T$ = 300, 340 and 350 K. Optical illumination is focused at the heterostructure with a laser at $\lambda$ = 900 nm, $P$ = 17 μW (see inset schematic). (c) $I_{ds}$-$V_{ds}$ curves (in logarithmic scale) with and without illumination. The wavelength of the laser is 4.034 μm, and the power density is 1.09 W cm$^{-2}$. Inset: Optical image of a typical b-AsP/MoS$_2$ heterostructure device. Scale bar, 5 μm. (d) Wavelength dependence of $D^*$ at $V_{ds}$ = 0 V. The purple and dark lines are commercial specific detectivity for a thermistor bolometer and PbSe MWIR detectors, respectively, at room temperature. (e) Schematic representation of photoexcited charge carrier dynamics in graphene-TMDs-graphene heterostructures in out-of-plane direction. Following pulsed-laser excitation, electron–hole pairs are created, separated and transported to the graphene electrodes. (f) Schematic band diagram for Gr/WS$_2$/Gr heterostructure with a built-in electric field to separate the generated e-h pairs. (g) Schematic representation of separation and transfer of photoexcited electrons and holes in Gr-MoS$_2$-Gr heterostructures. (h) External quantum efficiency in this device as a function of light power with different wavelength. (a,b) Reproduced with permission.[136] Copyright 2017, Nature Publishing Group. (c,d) Reproduced with permission.[140] Copyright 2017, American Association for the Advancement of Science. (e) Reproduced with permission.[144] Copyright 2016, Nature Publishing Group. (f) Reproduced with permission.[142] Copyright 2013, American Association for the Advancement of Science. (g,h) Reproduced with permission.[143] Copyright 2013, Nature Publishing Group.

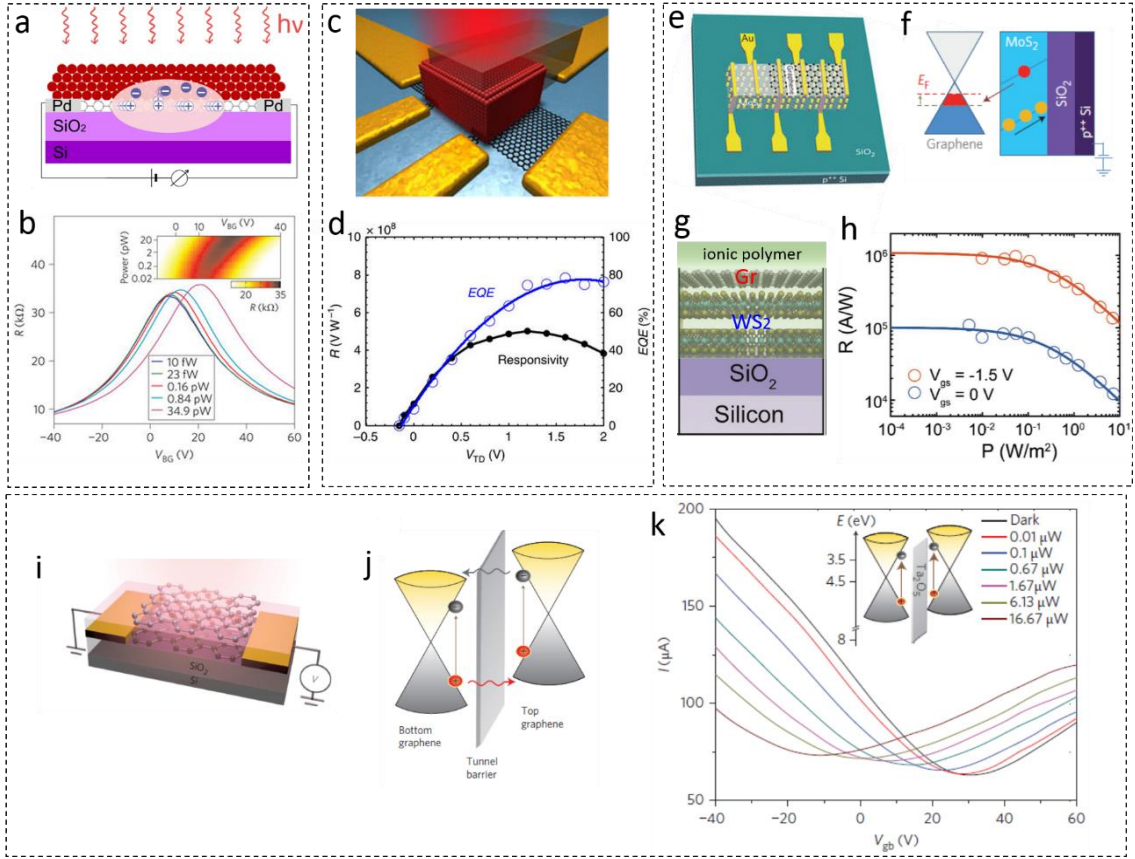

**Figure 5.** (a) Schematic of graphene-CQDs hybrid photodetectors, where graphene acts as carrier transport channel and CQDs act as strong light absorption layers. (b) Resistance as a function of back-gate voltage for the graphene–quantum dots structure for increasing illumination intensities. Increasing the illumination leads to a photogating effect that shifts the Dirac point to higher back-gate voltage due to the hole doping on graphene channel. Inset: two-dimensional plot of graphene resistance as a function of optical power. (c) Schematic of graphene-PbS QDs hybrid photodetector integrated with a top transparent electrodes. (d) Responsivity and EQE of the phototransistor as function of applied $V_{TD}$. (e) Schematic of device architecture based on graphene-$MoS_2$ heterostructures. (f) Schematic of charge exchange process for $V_g \ll V_T$ at interface between graphene and $MoS_2$. (g) Schematic diagram of $WS_2$-graphene photodetector with ionic polymer as top gate. (h) Responsivity of the device as a function of light power. (i) Schematic of phototransistors composed of a pair of stacked graphene layers sandwiching a thin tunnel barrier (5nm $Ta_2O_5$). (j) Schematic of band diagram and photoexcited hot carrier transport under light illumination. Vertical arrows represent photoexcitation, and lateral arrows represent tunnelling of hot electron (grey) and hole (red). (k) I-V characteristics of the device under different laser powers. Inset: Energy band diagram of the graphene/$Ta_2O_5$/graphene heterostructures. (a,b) Reproduced with permission.[147] Copyright 2012, Nature Publishing Group. (c,d) Reproduced with permission.[154] Copyright 2016, Nature Publishing Group. (e,f) Reproduced with permission.[162] Copyright 2013, Nature Publishing Group. (i-k) Reproduced with permission.[165] Copyright 2014, Nature Publishing Group.

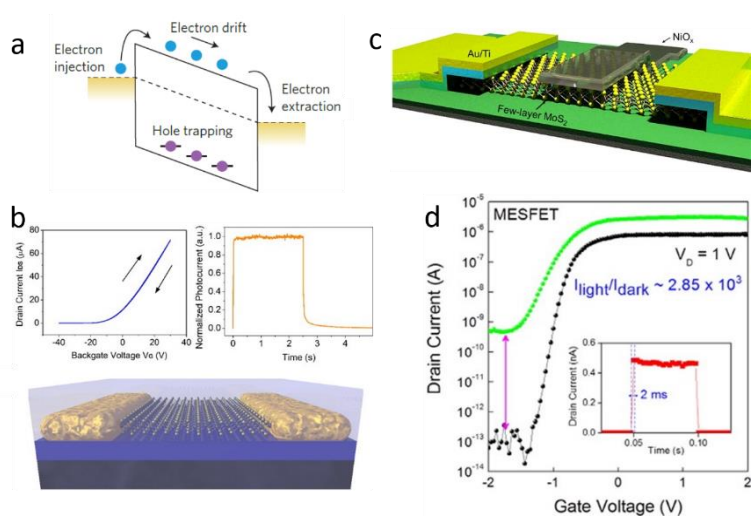

**Figure 6.** (a) Band diagram of a $MoS_2$ based photoconductive detector, taking into account hole trapping in the traps states closed to the valence band. (b)Transfer characteristics and photo-response dynamic of monolayer $MoS_2$ based phototransistors with $HfO_2$ encapsulation. (c) Schematic 3D view of $MoS_2$ based MESFET with 6 μm long channel, 3 μm long $NiO_x$ gate. (d) Transfer curves of the $MoS_2$ based MESFET under dark and green light illumination, Inset is temporal response of the devices. (a) Reproduced with permission.[100] Copyright 2014, Nature Publishing Group. (b) Reproduced with permission.[50] Copyright 2015, American Chemical Society. (c,d) Reproduced with permission.[177] Copyright 2015, American Chemical Society.

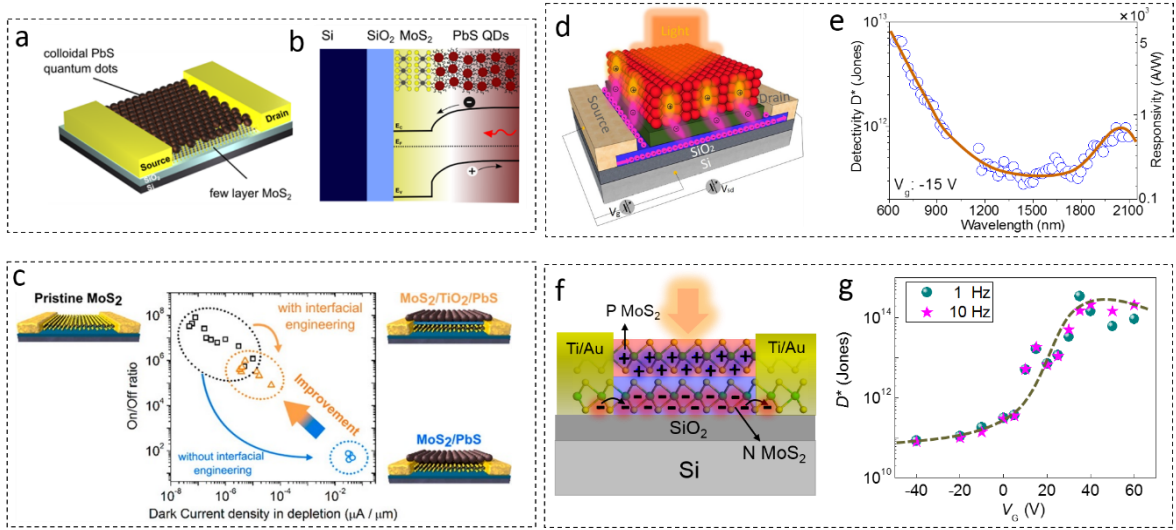

**Figure 7.** (a) Schematic of MoS$_2$-PbS QDs hybrid phototransistors. (b) Charge transfer at the inerface between MoS$_2$ and QDs under light illumination. (c) On/Off ratio versus dark current density in pristine MoS$_2$, MoS$_2$/PbS and MoS$_2$/TiO$_2$/PbS based phototransistors, the utilization of TiO$_2$ buffer layer at interface preserves the large On/Off ratio and low dark current after QDs sensitization. (d) Schematic diagram of MoS$_2$ and HgTe QDs hybrid based phototransistor with TiO$_2$ buffer layer. (e) Detectivity and responsivity spectral in the depletion mode. (f) Schematic diagram of MoS$_2$ phototransistors integrated with an out-of-plane p-n homojunction, where the built-in field in out-of-plane direction can facilitate the photoexcited carrier separation and lead to the photo-gating effect. (g) Detectivity of the detector as a function of back gate at bandwidth of 1Hz and 10Hz. (c) Reproduced with permission.[180] Copyright 2016, American Chemical Society. (f, g) Reproduced with permission.[188] Copyright 2017, Nature Publishing Group.

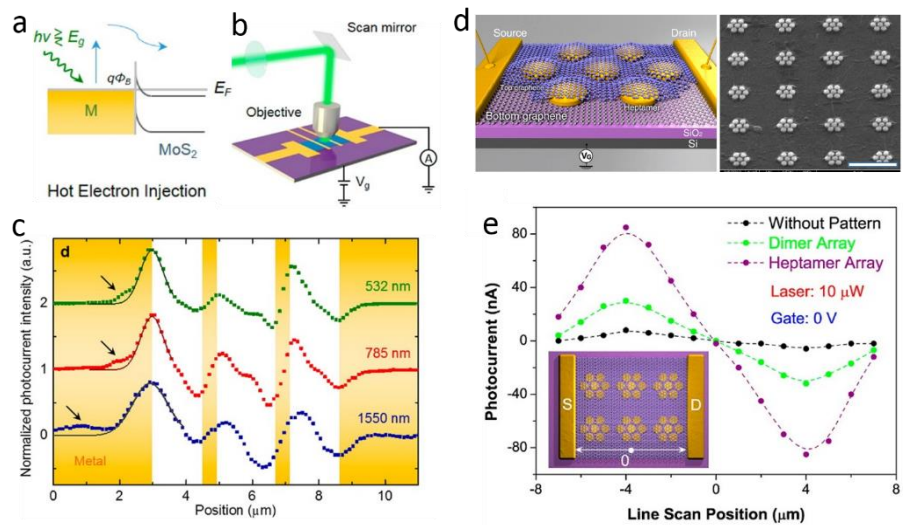

**Figure 8.** (a) Schematic illustration of hot electron injection from a metal electrode to MoS$_2$. $E_g$ represents the bandgap of MoS$_2$. (b) Schematic illustration of the MoS$_2$ device and the optical setup scanning across the MoS$_2$ channel. (c) Line profiles of the photocurrent response by scanning the laser beam along the channel with different wavelength. The orange background indicates electrode positions. The appeared photocurrent from NIR laser illumination is owing to the hot electrons from metal to MoS$_2$. (d) Schematic illustration (left panel) and SEM images (right panel) of gold heptamer array sandwiched between two monolayer graphene sheets. The scale bar is 1 µm. (e) Photocurrent measurements show antisymmetric photocurrent responses from the different regions of the device corresponding to specific plasmonic antenna geometries, obtained along the line scan direction in inset. (a-c) Reproduced with permission.[207] Copyright 2015, American Chemical Society. (d,e) Reproduced with permission.[200] Copyright 2012, American Chemical Society.

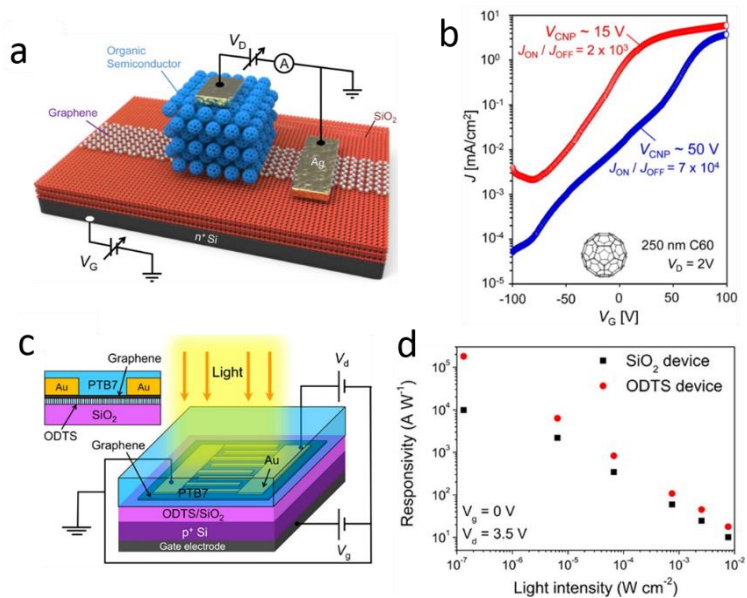

**Figure 9.** (a) Schematic illustration of graphene-organic semiconductor based vertical field effect transistors (VFETs). (b) Transfer characteristics of Graphene-$C_{60}$ VFETs. (c) Schematic illustration of a graphene-PTB7 hybrid photodetector. Inset shows the side view of the device. (d) Responsivity of the $SiO_2$ and ODTS devices with respect to light intensities. (a,b) Reproduced with permission.[220] Copyright 2015, American Chemical Society. (c,d) Reproduced with permission.[226] Copyright 2017, American Chemical Society.

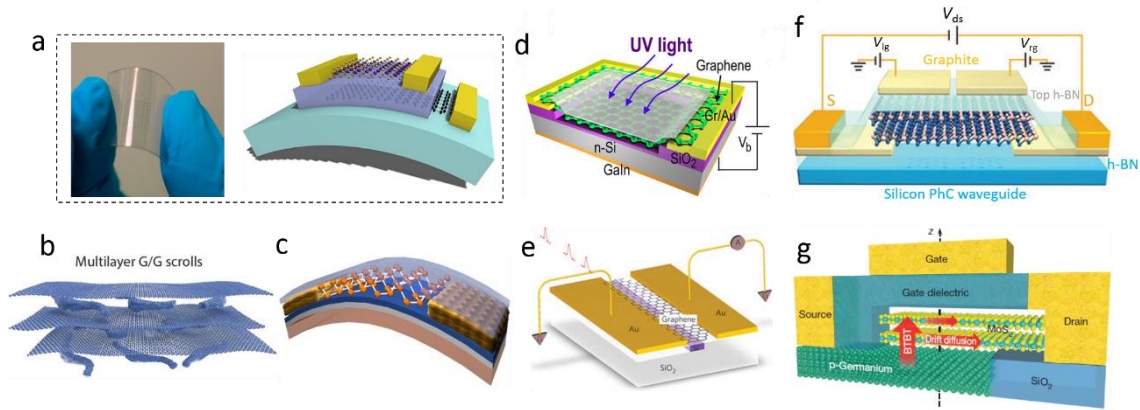

**Figure 10.** (a) Photograph (left panel) and schematic diagram (right panel) of all-2D based flexible photodetectors. (b) Schematic of intercalating graphene scrolls in between graphene layers. (c) Schematic of monolayer black phosphorus transistors on flexible substrate. (d) Schematic diagram of graphene-Si based UV photodetectors. (e) Schematic of the graphene photodetectors integrated with silicon waveguide. (f) Cross-sectional schematic of the encapsulated bilayer MoTe$_2$ p–n junction on top of a silicon PhC waveguide. (g) Schematic diagram illustrating the cross-sectional view of the tunnel field-effect transistors (TFET) composing of bilayer MoS$_2$ as the channel and degenerately doped p-type Ge as the source. (a) Left panel. Reproduced with permission.[251] Copyright 2014, American Chemical Society. Right panel. Reproduced with permission.[249] Copyright 2013, American Chemical Society. (b) Reproduced with permission.[255] Copyright 2017, American Association for the Advancement of Science. (c) Reproduced with permission.[256] Copyright 2015, American Chemical Society. (d) Reproduced with permission.[264] Copyright 2017, Nature Publishing Group. (e) Reproduced with permission.[269] Copyright 2013, Nature Publishing Group. (f) Reproduced with permission.[270] Copyright 2017, Nature Publishing Group. (g) Reproduced with permission.[271] Copyright 2015, Nature Publishing Group.

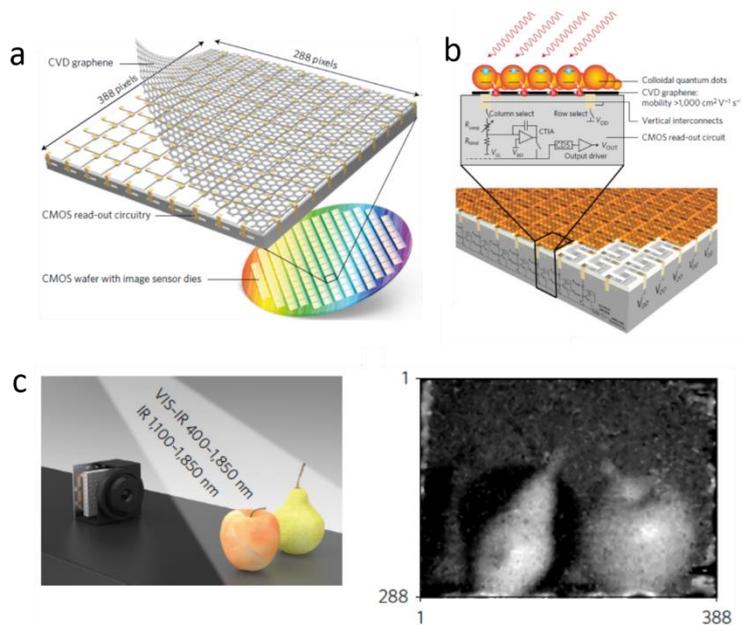

**Figure 11.** (a) Computer-rendered impression of the CVD graphene transfer process on a single die (real dimensions 15.1 mm height, 14.3 mm width) containing an image sensor read-out circuit that consists of 388 × 288 pixels. (b) Side view explaining the graphene phototransistor and the underlying read-out circuit. The bottom panel represents 3D impression of the monolithic image sensor displaying the top level with graphene carved into S-shaped channels sensitized with a layer of quantum dots, vertical interconnects and underlying CMOS read-out circuitry. (c) Digital camera set-up: the image sensor plus lens module captures the light reflected off objects that are illuminated by an external light source. The right panel is the near-infrared (NIR) and short-wave infrared (SWIR) light photograph of an apple and pear. Reproduced with permission.[274] Copyright 2017, Nature Publishing Group.

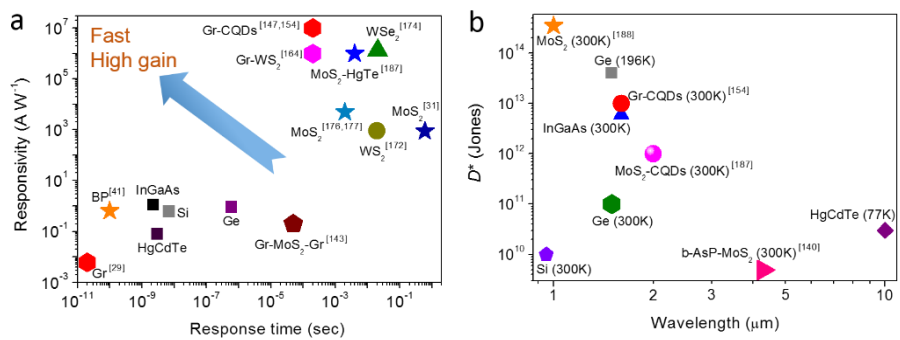

**Figure 12.** (a) Comparison of responsivity versus response time between commercial photodetectors and 2D or hybrid based photodetectors. (b) Specific detectivity $D^*$ versus response wavelength in conventional, 2D and hybrid photodetectors.

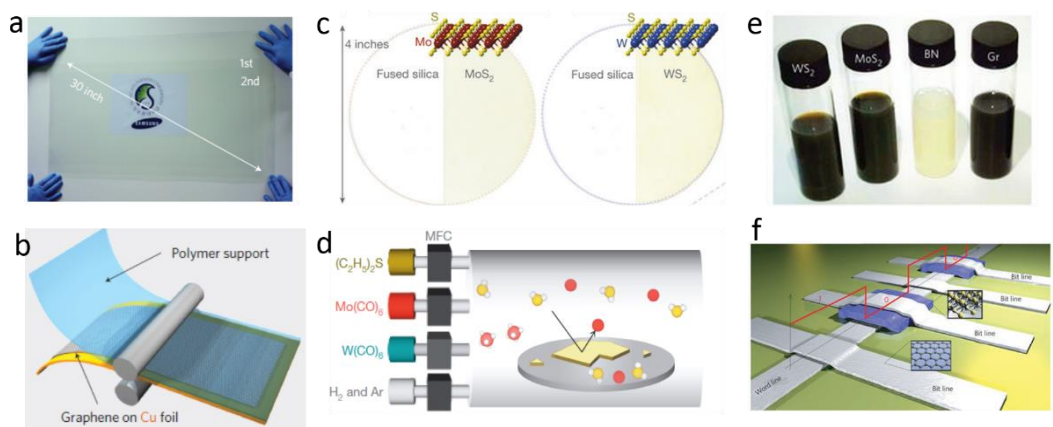

**Figure 13.** (a) A transparent ultra-large area graphene film transferred on a 35-inch PET sheet. (b) Schematic of the roll-based production of graphene films grown on a copper foil. (c) Photographs of monolayer $MoS_2$ and $WS_2$ films grown on 4-inch fused silica substrates, with diagrams of their respective atomic structures. The left halves show the bare fused silica substrate for comparison. (d) Diagram of MOCVD growth setup. Precursors were introduced to the growth setup with individual mass flow controllers (MFC). (e) Optical image of water-based two-dimensional crystal inks including 2D TMDs, h-BN and graphene. (f) Sketch of the fabricated logic memory devices using inkjet-printed method with 2D crystal inks. (a,b) Reproduced with permission.[312] Copyright 2010, Nature Publishing Group. (c,d) Reproduced with permission.[35] Copyright 2015, Nature Publishing Group. (e,f) Reproduced with permission.[83] Copyright 2017, Nature Publishing Group.